\documentclass[conference]{IEEEtran}
\usepackage{fancyhdr} 
\usepackage{cite}
\usepackage{amsmath,amssymb,amsfonts}
\usepackage{algorithmic}
\usepackage{multicol}
\usepackage{graphicx}
\usepackage{textcomp}
\usepackage{xcolor}
\usepackage{multirow}
\usepackage{hyperref}
\usepackage{booktabs}
\usepackage{sc24repro}
\def\BibTeX{{\rm B\kern-.05em{\sc i\kern-.025em b}\kern-.08em
    T\kern-.1667em\lower.7ex\hbox{E}\kern-.125emX}}
\usepackage{tikz}
\usetikzlibrary{positioning}

\newcommand{\ourlibrary}{pytket-cutensornet}
\newcommand{\bigO}[1]{\mathcal{O}(#1)}
\newcommand{\Rz}{\texttt{R}_\texttt{Z}}
\newcommand{\Rxx}{\texttt{R}_\texttt{XX}}
\newcommand{\SWAP}{\texttt{SWAP}}

\begin{document}

\bibliographystyle{IEEEtran}

\title{Realizing Quantum Kernel Models at Scale with Matrix Product State Simulation}

\author{
\IEEEauthorblockN{Mekena Metcalf}
\IEEEauthorblockA{\textit{Innovation and Ventures} \\
\textit{HSBC Holdings Plc.}\\
San Francisco, CA, USA \\
mekena.metcalf@us.hsbc.com}

\and
\IEEEauthorblockN{Pablo Andr\'es-Mart\'inez}
\IEEEauthorblockA{\textit{Quantinuum}\\
Cambridge, United Kingdom \\
pablo.andresmartinez@quantinuum.com}

\and
\IEEEauthorblockN{Nathan Fitzpatrick}
\IEEEauthorblockA{\textit{Quantinuum}\\
Cambridge, United Kingdom \\
nathan.fitzpatrick@quantinuum.com}
}

\maketitle

\begin{abstract}
Data representation in quantum state space offers an alternative function space for machine learning tasks. However, benchmarking these algorithms at a practical scale has been limited by ineffective simulation methods. We develop a quantum kernel framework using a Matrix Product State (MPS) simulator and employ it to perform a classification task with 165 features and 6400 training data points, well beyond the scale of any prior work. We make use of a circuit ansatz on a linear chain of qubits with increasing interaction distance between qubits. We assess the MPS simulator performance on CPUs and GPUs and, by systematically increasing the qubit interaction distance, we identify a crossover point beyond which the GPU implementation runs faster. We show that quantum kernel model performance improves as the feature dimension and training data increases, which is the first evidence of quantum model performance at scale.
\end{abstract}


\section{Introduction}
Representation of classical data in quantum state space with near-term quantum algorithms is anticipated to improve existing learning tasks like classification or time-series analysis ~\cite{SchuldBook}. The intersection of quantum information and machine learning sparked considerable engagement from industries seeking to apply these new algorithms to enhance existing machine learning frameworks~\cite{magnusson2022,metcalf2023,PVforecast,creditscore2023, QMLDrug}. Much of this testing uses toy models with unrealistic data specifications for any genuine integration into an industrial, production machine learning environment, largely due to algorithm scaling difficulty. This paper and its associated application framework address the obstacles to testing quantum machine learning algorithms at scale, specifically for hybrid quantum algorithms that use a kernel method. Previous works benchmarking quantum kernel methods were both limited in feature dimension and limited in data dimension, leading to mixed results regarding the quantum kernel effectiveness compared to Gaussian kernels (among other methods) for classification tasks~\cite{Bowles2024, Slattery2023, Egginger2023}. The proposed simulation framework enables quantum machine learning practitioners to push well beyond previous limitations. 

There is theoretical evidence that quantum feature map expressiveness improves model generalization~\cite{Schuld2021,Peters2023generalization, Huang2021}, though quantum model generalization capability receives continued debate in the literature for quantum neural networks~\cite{Gil-Fuster2024}. Model performance assessment on classical data sets is largely determined through numerical benchmarking efforts, yet vector representations of quantum states require an amount of memory that scales exponentially with the number of qubits, making simulation at meaningful scale prohibitive, particularly for feature maps encoding classical data with quantum Hamiltonians. Alternatively, one may employ a quantum computer to test these quantum machine learning algorithms at scale; however, noise arising from imperfect operations and other environmental effects degrade learning performance by contributing to exponential concentration in the case of kernel algorithms and barren plateaus in the case of variational quantum algorithms~\cite{Wang2021, ExpConc}. This inability to benchmark quantum machine learning algorithms at scale in both feature dimension and training data size leads to inconsistency in the literature about whether quantum machine learning is an effective application area compared to classical learning algorithms. 

Quantum circuit simulation using tensor networks has progressed in tandem with the scaling of quantum computers, often acting as the classical framework competing against quantum advantage claims~\cite{Gray2021,Tindall2024, Begusic2024}. The scaling bottleneck of tensor network methods lies in the number of two-qubit quantum circuit operations (gates) rather than the number of qubits. This makes tensor network simulators effective for low-depth quantum circuits and enables the use of approximate methods to reach a scale of 100 or more qubits~\cite{Ayral2023}. Further, tensor network simulators yield the full mathematical description of the quantum state, which can be used directly to calculate the similarity measure---inner product---between states necessary for the quantum kernel method used in this manuscript. Alternatively, one may use the projected quantum kernel approach to compute an observable set for each data point and get the kernel element classically~\cite{Huang2021}. 

Recently Ref~\cite{Chen2024} proposed an alternative GPU accelerated tensor network framework for the quantum support vector machine. While the motivation is similar to our proposed methodology, the implementation and analysis are distinct. Their paper encodes data using a block-encoded state ansatz where complexity is increased with increased block dimensionality, and only a linear connectivity is considered in the analysis. Our work considers encoding data through a spin Hamiltonian with a specified interaction distance, and we further assess the computational resources needed with increased complexity along with the resulting model performance effect. Our approach breaks apart the problem so that the number of MPS simulations scales linearly with the number of data points, with the quadratic scaling only applying to inner product calculations, which are relatively inexpensive. In Ref~\cite{Chen2024}, the number of tensor network simulations scales quadratically with the number of data points. Another major distinction from Ref~\cite{Chen2024} is that we make use of tensor network truncation methods to significantly reduce the simulation overhead, while guaranteeing the corresponding truncation error does not impact the results.
Ref~\cite{Chen2024} shows GPU performance gains over CPUs on their tensor network contraction task, but their paper doesn't consider analysis on the tensor network complexity and the genuine need of GPUs to reach a practical scale. 
The simulation method used in Ref~\cite{Chen2024} is based on tensor network contraction via contraction path optimisation. They re-use their contraction path for subsequent simulations, which is possible since the circuit topology remains fixed while parameters or data points are changed. Contraction path optimization methods are currently not well suited for the truncated MPS methods we employ, since it becomes difficult to guarantee truncation error~\cite{Gray2024}. While their methods are designed to solve the same problem, our approach provides additional analysis with the ultimate conclusion that CPUs suffice to construct quantum kernels for the support vector classifier.

This paper introduces a tensor network framework designed for quantum kernel methods and demonstrates tensor network effectiveness at scaling this application. Quantum kernels capture the distance between data points in quantum feature space by evaluating the quantum state overlaps associated with each data point. Expressing data in quantum feature space is postulated to produce more separable data that improves the results of linear classifiers like the Support Vector Machine~\cite{Havlicek2019}. The different quantum state overlaps may be computed independently, and we exploit parallel processing to significantly reduce computational time, enabling us to train on more data. We show that quantum kernels continue to improve classification metrics with the addition of more training data and more features.

We leverage GPU computing architectures and analyze whether they provide a runtime advantage for tensor network simulations in our application-specific framework. To achieve this, we introduce \ourlibrary{}~\cite{pytket-cutensornet}, a new library for quantum circuit simulation on GPUs using tensor networks. We conduct benchmarking of GPU simulations against a state of art CPU implementation facilitated by the ITensors software library~\cite{itensor}, employing quantum circuits of varying complexity. Specifically, we systematically increase the interaction distance between qubits in a linear chain, thereby augmenting qubit correlation and increasing quantum feature map expressivity. Our analysis reveals a crossover point in runtime between CPU and GPU implementations as the circuit complexity reaches a specific threshold. 

\section{Quantum Kernel Method via MPS}

We describe the mathematical framework for constructing quantum kernels to evaluate data similarity within a quantum Hilbert space. Our approach employs quantum feature maps, which translate classical data into the coefficients of a spin Hamiltonian and are represented as quantum circuits through Hamiltonian exponentiation. The number of qubits in our circuits match the number of features in the data set. We use tensor networks to simulate these circuits, with computational complexity primarily determined by the inclusion of two-qubit gates rather than the number of qubits. To manage computational resources, we employ Singular Value Decomposition (SVD) to reduce the dimension of tensors while guaranteeing an accuracy threshold. In our quantum kernel application, we simulate a quantum circuit for each individual data point in the data set and compute the overlap between pairs of data points by calculating the inner product between their respective quantum states. These computations are performed independently for each data point, allowing us to leverage parallelization techniques. To assess performance across different platforms and code bases, we evaluate our computational framework using ITensors~\cite{itensor} on CPUs and our own implementation of MPS simulation methods on GPUs, powered by NVIDIA's cuTensorNet~\cite{cutensornet}.

\subsection{Quantum Kernel Method}
\label{sec:qkernel}

Kernel machines are used in a variety of machine learning methods to capture non-linearity in data. Support Vector machine (SVM) is the most well known linear classifier with a kernel capturing data non-linearity. Consider data existing in some space $\mathcal{D}$ and an additional finite-dimensional kernel Hilbert space $\mathcal{H}$ where there exists a map $\psi \colon \mathcal{D}\rightarrow \mathcal{H}$. Data is expressed in the kernel Hilbert space by applying the feature map $\psi(\boldsymbol{x})$ to each data point $\boldsymbol{x}$, and data similarity in the kernel Hilbert space is determined from the inner product $k(\boldsymbol{x},\boldsymbol{x^\prime}) = \langle \psi(\boldsymbol{x}),\psi(\boldsymbol{x^\prime})\rangle$. This inner product, known as the kernel function, makes up the Gram matrix with entries
\begin{equation} \label{eq:Gram}
\boldsymbol{K}_{ij} = \vert\langle \psi(\boldsymbol{x}_i),\psi( \boldsymbol{x}_j)\rangle\vert^2
\end{equation}
that captures similarity between N data points in the kernel Hilbert space. Expression of data in quantum Hilbert space can potentially better capture data non-linearity and improve machine learning performance~\cite{Peters2023generalization}.

A data vector $\boldsymbol{x}$ with $m$ features such that $\boldsymbol{x} = \left(x_1,..., x_m \right)$ is first rescaled to values in the $(0,2)$ real interval and then mapped to a quantum state on $m$ qubits $\lvert\psi(\boldsymbol{x})\rangle$ using a parameterized unitary operator $U(\boldsymbol{x})$ such that 
\begin{equation}
\lvert \psi(\boldsymbol{x}) \rangle = U(\boldsymbol{x})\lvert +\rangle^{m}.
\end{equation}
The state is first initialised in a uniform superposition $\lvert + \rangle^{m}$ by the application of a Hadamard gate on each of the $m$ qubits of a $\lvert 0 \rangle$ state. 
Then, the unitary operator $U(\boldsymbol{x})$ rotates the state within the Hilbert space differently for each data point $\boldsymbol{x}$.

Our unitary operator $U(\boldsymbol{x})$ is defined as a sequence of exponential operators
\begin{equation} \label{eq:U}
    U(\boldsymbol{x}) = \left( e^{-iH_{X\!X}(\boldsymbol{x})} \cdot e^{-iH_Z(\boldsymbol{x})} \right)^r
\end{equation}
where $r$ is a tunable parameter that determines the number of layers in the circuit, and $H_Z$, $H_{X\!X}$ are Hamiltonians describing single-qubit and two-qubit interactions:
\begin{align}
    H_Z(\boldsymbol{x}) &= \gamma\sum_{i=1}^m x_i \vec{\sigma}^Z_i \label{eq:Hz} \\
    H_{X\!X}(\boldsymbol{x}) &=  \gamma^2\,\frac{\pi}{2} \sum_{(i,j) \in G} (1\!-\!x_i)(1\!-\!x_j)\vec{\sigma}_i^X\vec{\sigma}_j^X. \label{eq:Hxx}
\end{align}
Thus, the unitary operator $U(\boldsymbol{x})$ resembles a Trotterized evolution of an Ising spin model Hamiltonian.
The qubit interaction topology in $H_{X\!X}$ is captured by the edges of a graph $G$; in our experiments this corresponds to a linear chain with tunable interaction distance (see section~\ref{sec:ansatze}). Additional edges in the graph contribute to more generators that define the underlying Lie algebra encoding the data, which increases the feature map expressivity.
The real coefficient, $\gamma$, controls the kernel bandwidth and is needed to scale to larger quantum feature spaces\cite{Shaydulin2022, Canatar2022}.
Our choice of feature map is similar to the original proposed in Ref.~\cite{Havlicek2019} and the Pauli operators have been chosen empirically based on model performance.

\subsection{Quantum Circuit Simulation}
\label{sec:simulation}

The state of an $m$-qubit quantum computer can be described by a vector of $2^m$ complex entries. 
We can simulate a quantum computation on a classical computer by storing such a state vector in memory and applying matrix multiplications to it for each gate in the circuit.
Since the number of entries in the vector scales exponentially with the number of qubits, simulation of circuits with more than $30$ qubits becomes prohibitively expensive. 
Previous works have pushed the limits of state vector simulators, with the recent simulation of a $40$ qubit circuit implementing Shor's algorithm being at the limit of what is feasible with current supercomputers \cite{ShorSimulated}.
The latter experiment required the use of all of the 2048 GPUs of a cluster with 40 GiB RAM per GPU to simulate a single circuit.

An alternative widely used approach across physics and chemistry is to represent the state of the computation as a tensor network, rather than a vector~\cite{MPS_sim, Gray2021, Tindall2024}.
Such methods do still suffer from exponential scaling, but instead of it being with respect to the number of qubits, the scaling depends on the amount of entanglement that is built up during the computation.
Consequently, circuits with large numbers of qubits but only a few layers of gates can be efficiently simulated: a recent example being the simulation of a Trotterized kicked Ising experiment on a heavy-hex lattice of $127$ qubits~\cite{Tindall2024}.

Tensors are multidimensional arrays on which we define an operation akin to matrix multiplication.
They are often presented in diagrammatic notation, where a tensor with three axis corresponds to a node with three `legs':
\begin{center}
    \begin{tikzpicture}
        \node (picture) {
        \begin{tikzpicture}
            \draw (0,0) circle (0.2);
            \draw (0,0.2) -- (0,0.6) node[near end,left] {\footnotesize $j$\!};
            \draw (0.2,0) -- (0.7,0) node[near end,above] {\footnotesize $k$\!};
            \draw (-0.2,0) -- (-0.7,0) node[near end,above] {\footnotesize \!$i$};
        \end{tikzpicture}
        };
        \node[right=7mm of picture] (array) {$T[i][j][k]$};
    \end{tikzpicture}
\end{center}
We refer to each of the axes of the array as a bond.
The range of possible indices in each axis determines its \textit{bond dimension}.
A standard matrix is a tensor with only two bonds, and the total number of entries in a tensor can be calculated as the product of all bond dimensions.

Two tensors can be \textit{contracted} together along a common bond.
The entries of the resulting tensor are calculated from the equation
\begin{equation}
    C_{abxyz} = \sum_{s=0}^{\chi_s-1} A_{abs} \cdot B_{sxyz}
\end{equation}
where $A_{abs}$ is a shorthand for the array entry $A[a][b][s]$ and $\chi_s$ is the dimension of the bond with $s$ indices.
Moreover, we can rearrange the entries of a tensor into another tensor with a different number of bonds.
For instance, we can reshape any tensor into a standard matrix:
\begin{equation}
    M[i][j] = C_{abxyz} \quad \text{where} \quad 
    \begin{aligned}
        i&=a+b\!\cdot\!\chi_a \\
        j&=x+(y+z\!\cdot\!\chi_y)\chi_x
    \end{aligned}
\end{equation}
where the way we define $i$ and $j$ can be any arbitrary bijection.\footnote{We simply need to establish a bijection $[\chi_a] \times [\chi_b] \cong [\chi_a\chi_b]$ for the bond indices $(a,b) \mapsto i$, and similarly for $(x,y,z)$ so that we keep track of where each of the entries from $C$ are located in $M$. Here, $[n]$ stands for the range of integers from $0$ to $n-1$.}
Thus, standard matrix decompositions such as SVD are also available to tensors by first reshaping them into matrices.

We can connect multiple tensors together to form a tensor network.
A Matrix Product State (MPS) is a standard instance of a tensor network whose structure is presented below for the case of $5$ qubits.
\begin{center}
    \begin{tikzpicture}
        \draw (0,0) circle (0.2);
        \draw (0,0.2) -- (0,0.5);
        \draw (1,0) circle (0.2);
        \draw (1,0.2) -- (1,0.5);
        \draw (-1,0) circle (0.2);
        \draw (-1,0.2) -- (-1,0.5);
        \draw (2,0) circle (0.2);
        \draw (2,0.2) -- (2,0.5);
        \draw (-2,0) circle (0.2);
        \draw (-2,0.2) -- (-2,0.5);
        \draw (0.2,0) -- (0.8,0);
        \draw (-0.2,0) -- (-0.8,0);
        \draw (1.2,0) -- (1.8,0);
        \draw (-1.2,0) -- (-1.8,0);
    \end{tikzpicture}
\end{center}
Bonds connecting tensors together are known as \textit{virtual bonds} and their dimension changes throughout the simulation.
Each of the open bonds correspond to one of the qubits in the quantum state and their dimension is fixed to $2$.
Pairwise contracting all of the tensors in an MPS along their common virtual bonds results in a single tensor with $m$ bonds of dimension $2$ and, hence, $2^m$ entries.
Thus, we can retrieve the state vector representation of a state from its MPS.

There is an advantage to representing a quantum state on $m$ qubits as an MPS when the dimension of the virtual bonds is upper bounded by some relatively small constant $\chi$.
Then, the total number of entries in the MPS is $2m\chi^2$ which will be considerably smaller than $2^m$ when $\chi \ll 2^{m/2}$.
As such, representing the quantum state under simulation as an MPS has the potential of saving a considerable amount of memory, allowing us to perform simulations on a large number of qubits.
Moreover, the time complexity of tensor contraction and decomposition scales as a low degree polynomial of the bond dimension and, hence, a small value of $\chi$ also enables fast simulation.
Crucially, the bond dimension depends on the strength of the entanglement present in the quantum state, meaning only quantum states with low entanglement are efficiently representable as an MPS.

\begin{figure}
    \centering
    \begin{tikzpicture}
        \node (rz) {\includegraphics[scale=0.55]{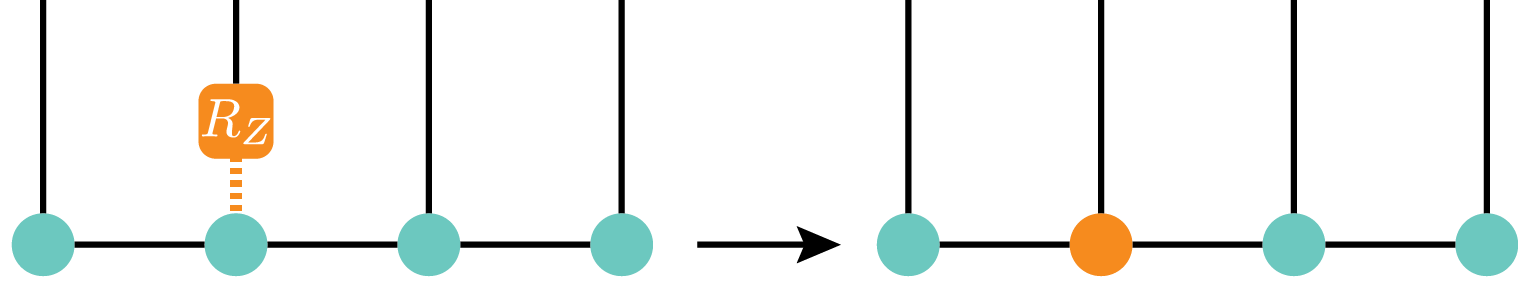}};
        \node[below=8mm of rz.south] (rxx) {\includegraphics[scale=0.55]{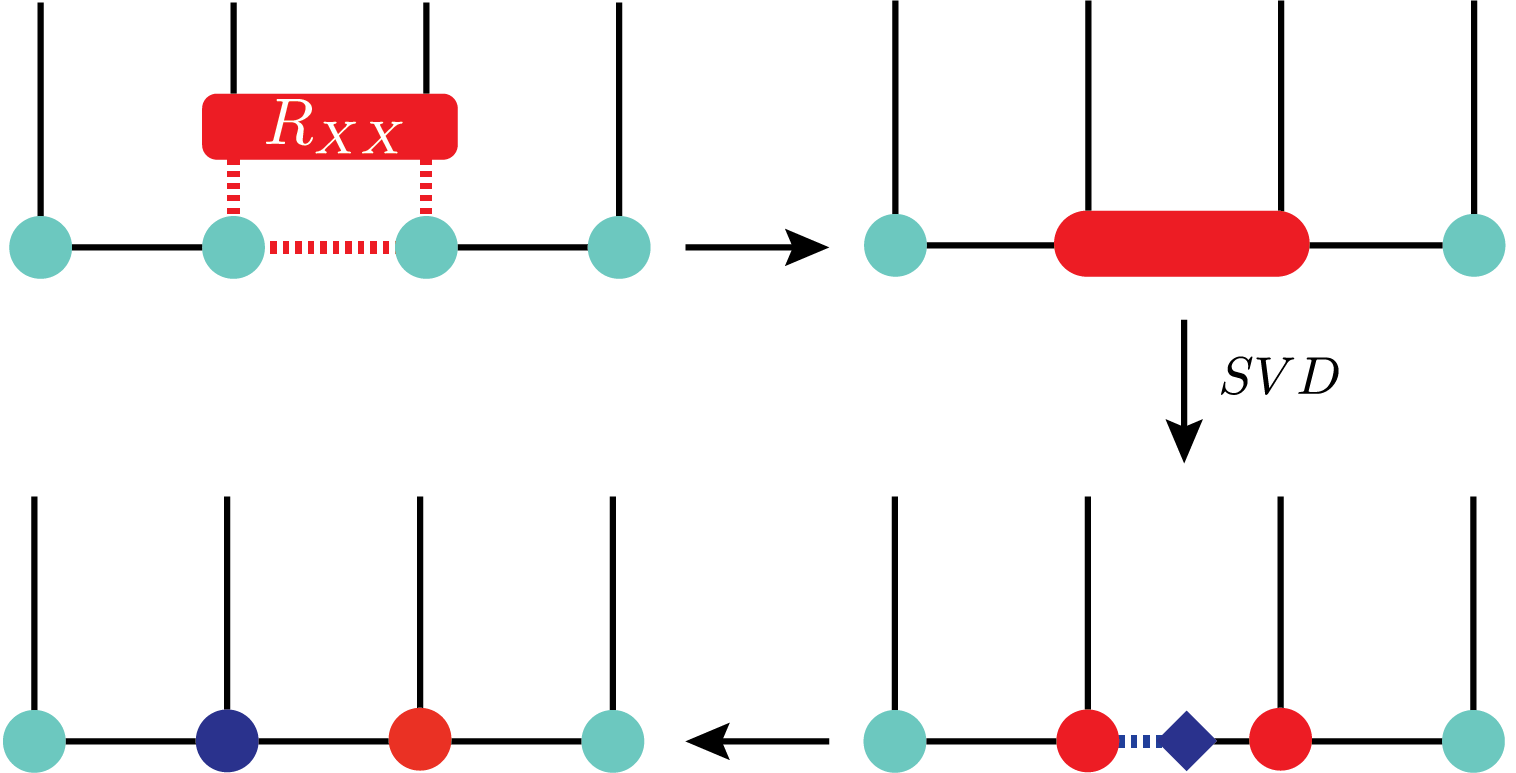}};
        \node[above left=-2mm and 0mm of rz.north west] (a) {(a)};
        \node[above left=-2mm and 0mm of rxx.north west] (b) {(b)};
    \end{tikzpicture}
    \caption{Gate application on a state represented as an MPS (turquoise) where bonds to be contracted are represented by dotted lines. (a) Single-qubit gates are contracted with the corresponding site tensor. (b) Two-qubit gates are applied in three steps: contract with the two MPS tensors, apply SVD decomposition on the result and finally contract the diagonal tensor of singular values (diamond) with either of the other tensors resulting from the decomposition.}
    \label{fig:gate_application}
\end{figure}

Fig.~\ref{fig:gate_application} depicts the steps required to update an MPS with the application of a quantum gate~\cite{MPS_sim}.
Two-qubit gates can only be applied on qubits that are adjacent in the MPS representation of the state. 
As discussed in section~\ref{sec:ansatze}, circuits that require richer connectivity are preprocessed to satisfy this constraint. 
Whenever a two-qubit gate is applied, the virtual bond between the qubits it acts on increases in dimension by a multiplicative factor.
Consequently, the virtual bond dimension increases exponentially with the number of two-qubit gates applied to it.
Unlike in state vector simulators, our bottleneck is not the number of qubits in the circuit, but the number of two-qubit gates.
This trade-off is shared by all simulation approaches based on tensor network representations of the state~\cite{MPS_sim, Gray2021, Tindall2024}.

The dimension of virtual bonds can be reduced during simulation if approximations are allowed.
This is realized when applying singular value decomposition (SVD) during two-qubit gate application (Fig.~\ref{fig:gate_application}b).
SVD factorizes the entries of a tensor $T$ into a sum $T_{xyzw} = \sum_i s_i A_{xyi} B_{izw}$ where $A$ and $B$ are tensors and there is a real number $s_i$ for each $i$, known as the \textit{singular value}.
When $s_i \approx 0$ for any given $i$, the contribution of entries $A_{xyi}$ and $B_{izw}$ becomes negligible, so we can remove them from the tensor to reduce the dimension of the bond indexed by $i$.
This process is known as SVD \textit{truncation} and it is the basis of approximate tensor network simulation methods~\cite{MPS_sim, Tindall2024}.

Assuming the MPS is normalized and in canonical form,\footnote{Standard MPS simulation algorithms apply a global transformation on the MPS known as \textit{canonicalization} which guarantees that the truncation is optimal in terms of accuracy~\cite{MPS_intro}. Canonicalization is applied before each SVD truncation.} 
we can quantify the error introduced by a single SVD truncation as: 
\begin{equation} \label{eq:trunc_error}
    |\langle \,\psi_{ideal},\, \psi_{trunc}\, \rangle |^2 \ =\ 1 - \sum s_i^2
\end{equation}
where the sum ranges over the truncated singular values.
Considering that errors due to 64-bit float point precision are at the scale of $10^{-16}$, in our experiments we choose to truncate singular values up to the point where  $\sum s_i^2 \sim 10^{-16}$.
Thus, we argue that the approximation errors we introduce with SVD truncation are at the scale of 64-bit machine precision and are essentially negligible.

\begin{figure*}
    \centering
    \begin{tikzpicture}
        \node (contraction) {\includegraphics[scale=0.63]{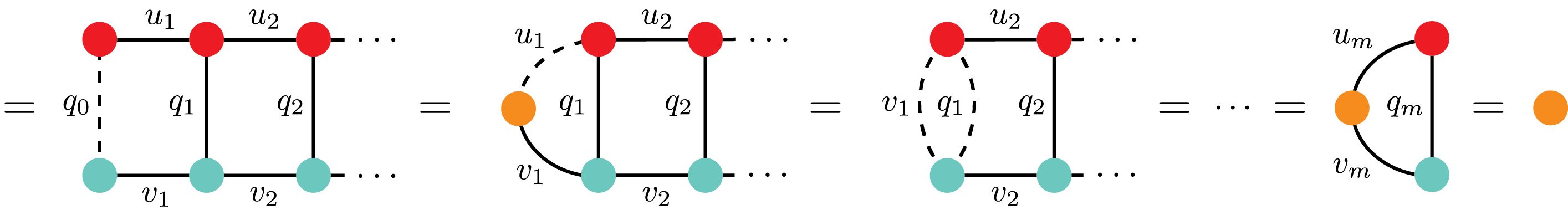}};
        \node[left=0mm of contraction.west] (expression) {$\langle \psi(\boldsymbol{x}_i), \psi(\boldsymbol{x}_j) \rangle$};
    \end{tikzpicture}
    \caption{Calculation of an inner product between states $\lvert \psi(\boldsymbol{x}_i) \rangle$ and $\lvert \psi(\boldsymbol{x}_j) \rangle$ in an MPS representation. All tensors of the (red) MPS representing $\lvert \psi(\boldsymbol{x}_i) \rangle$ have their entries conjugated. We connect both MPS by their qubit bonds and contract the resulting network in the order shown; dashed lines indicate where the contraction takes place at each step. The result is a tensor with no bonds: the complex number resulting from the inner product.}
    \label{fig:MPS_product}
\end{figure*}

Quantum kernel methods require us to calculate the inner products of pairs of quantum states.
Since our states are represented in MPS form, we need an efficient algorithm to contract a tensor network of two MPS to obtain the resulting inner product.
The tensor network to be contracted is depicted in Fig.~\ref{fig:MPS_product}, along with the order in which the contraction operations are applied.
This order leads to a time complexity of $\bigO{m\chi^3}$ for $m$ qubits and bond dimension $\chi$~\cite{MPS_Orus}.

Our quantum kernel framework is implemented in Python and is publicly available in GitHub.
We offer two backends for MPS simulation and calculation of inner products: ITensors (CPU) and our custom \ourlibrary{} library built upon cuTensorNet (GPU).
Both libraries use the same MPS simulation algorithm described in this section, and both are open-sourced.
ITensors~\cite{itensor} is a widely used library for tensor network simulations with focus on high-performance.
Our custom library makes use of NVIDIA's cuTensorNet~\cite{cutensornet} tensor contraction and decomposition primitives to run tensor network simulations on GPUs.
Other alternatives for MPS simulation on GPUs are cuTensorNet itself---which in a recent update included high-level API for MPS simulation, different from our own---and ITensors using a GPU backend.

\subsection{Circuit ansatz}
\label{sec:ansatze}

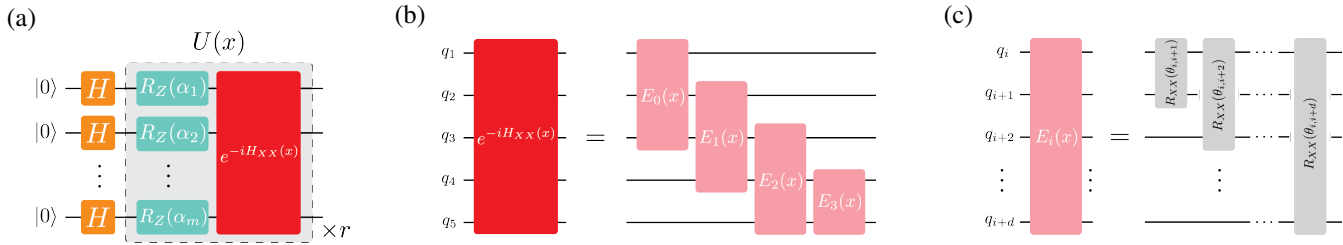
\begin{figure*}
    \centering
        \begin{tikzpicture}
        \node (circ_a) {\includegraphics[scale=0.57]{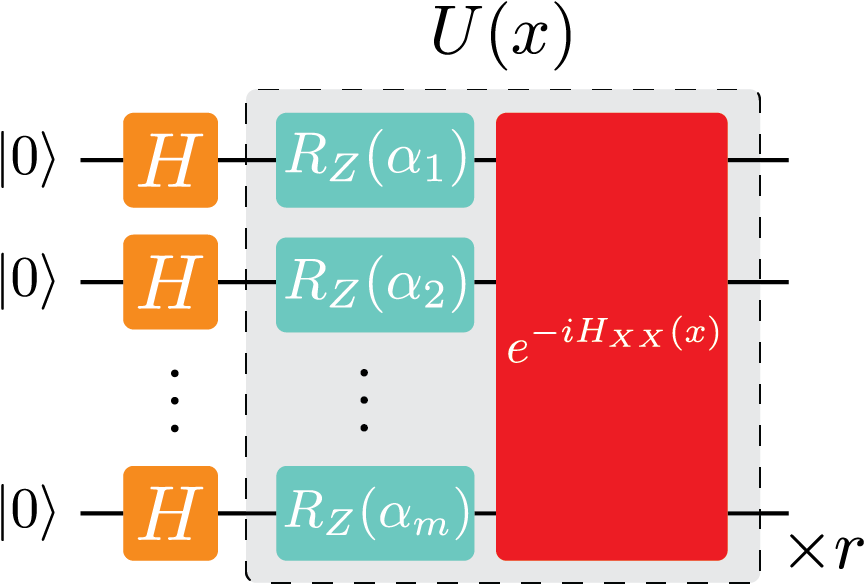}};
        \node[right=9mm of circ_a.east] (circ_b) {\includegraphics[scale=0.57]{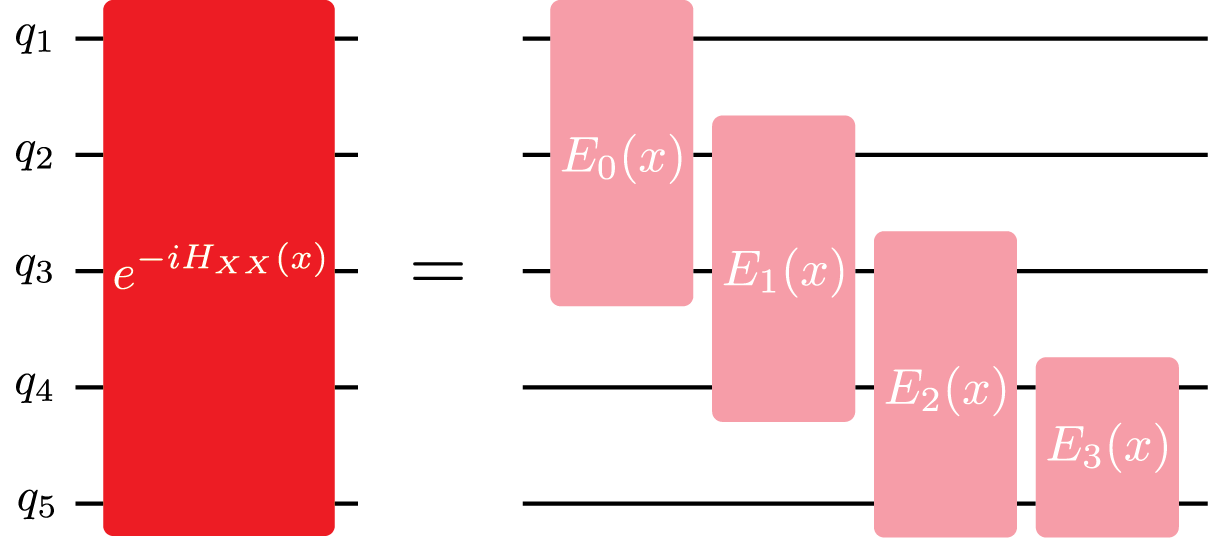}};
        \node[right=12mm of circ_b.east] (circ_c) {\includegraphics[scale=0.57]{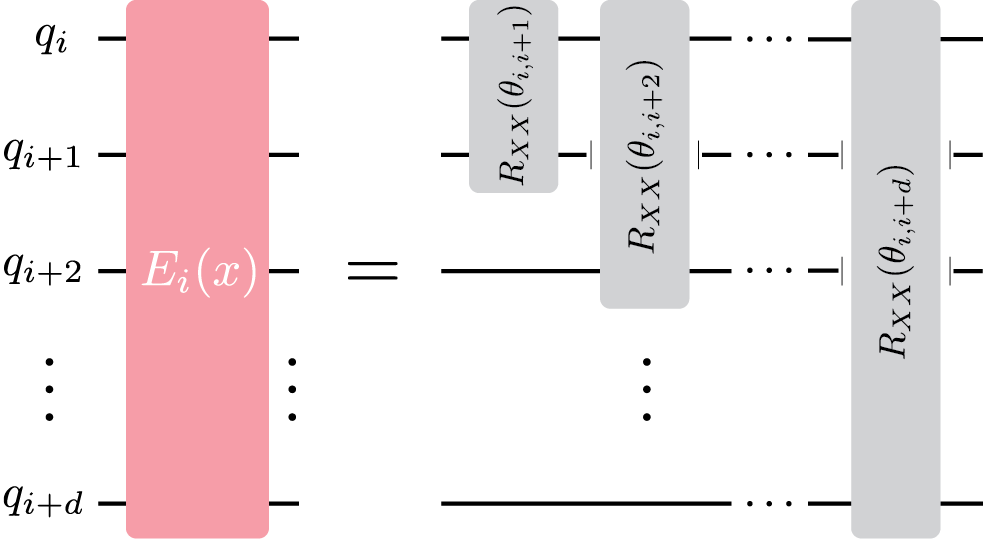}};
        \node[above left=-2.5mm and -2.5mm of circ_a.north west] (a) {(a)};
        \node[above left=-1mm and -1mm of circ_b.north west] (b) {(b)};
        \node[above left=-1mm and -1mm of circ_c.north west] (c) {(c)};
    \end{tikzpicture}
    \caption{The circuit ansatz preparing the state $U(\boldsymbol{x})\lvert + \rangle^m$ for each data point $\boldsymbol{x}$ of $m$ features. Circuits have $m$ qubits and a tunable number of layers ($r$) and interaction distance ($d$). (a) Overall circuit structure; the parameter $r$ indicates how many copies of $e^{-iH_{X\!X}(\boldsymbol{x})} \cdot e^{-iH_Z(\boldsymbol{x})}$ are applied; the angles of the $\Rz$ gates correspond to the coefficients in equation~\eqref{eq:Hz}. (b) The structure of the circuit for $e^{-iH_{X\!X}(\boldsymbol{x})}$ on $m=5$ qubits and interaction distance $d=2$. (c) The $\Rxx$ gates comprising a single $E_i(\boldsymbol{x})$ block from the $e^{-iH_{X\!X}(\boldsymbol{x})}$ operator; the angles of the $\Rxx$ gates correspond to the coefficients in equation~\eqref{eq:Hxx}.}
    \label{fig:ansatze}
\end{figure*}

In section~\ref{sec:qkernel} we introduced the Hamiltonians~\eqref{eq:Hz} and~\eqref{eq:Hxx} characterising the action of $U(\boldsymbol{x})$.
Fig.~\ref{fig:ansatze} depicts the parameterized circuit we use to generate each of the required $\lvert \psi(\boldsymbol{x}) \rangle = U(\boldsymbol{x}) \lvert + \rangle^m$ states.
The angles of the $\Rz$ and $\Rxx$ gates are determined by the feature values of the datapoint $\boldsymbol{x}$ and the kernel bandwidth $\gamma$, according to the coefficients of $H_{Z}(\boldsymbol{x})$ and $H_{X\!X}(\boldsymbol{x})$.
The two-qubit interaction topology we use in our experiments is depicted in Fig.~\ref{fig:ansatze}; namely, a linear chain of qubits with nearest neighbor interactions up to a tunable distance parameter $d$.
The arrangement of $\Rxx$ gates in Fig.~\ref{fig:ansatze} is chosen for the sake of clarity; in practice we take advantage of the fact that $\Rxx$ gates commute with each other to rearrange all of the gates within the subcircuit implementing $e^{-iH_{X\!X}(\boldsymbol{x})}$ to minimise circuit depth.\footnote{
It is straightforward to rearrange the $\Rxx$ gates so that all qubits have a gate applied to them at each time step and, since each qubit interacts with up to $2d$ qubits, it follows that the subcircuit implementing $e^{-iH_{X\!X}(\boldsymbol{x})}$ can be realised in $2d$ layers.}

Since the MPS simulation method we use only supports the application of two-qubit gates between adjacent qubits, only circuits with qubit interaction distance $d=1$ can be natively simulated.
For $d > 1$ we must add the necessary $\SWAP$ gates before each $\Rxx$ gate to bring the two qubits it acts on together.
After the $\Rxx$ gate is applied, the reverse sequence of $\SWAP$ gates is applied to return the qubits to their original position.
Thus, an $\Rxx$ gate acting on qubits at positions $i$ and $i+k$ in the linear chain requires an additional $2(k-1)$ $\SWAP$ gates.

\subsection{Parallelization}
\label{sec:parallel}

\begin{figure*}
    \centering
    \begin{tikzpicture}
        \node (nm) {\includegraphics[scale=0.33]{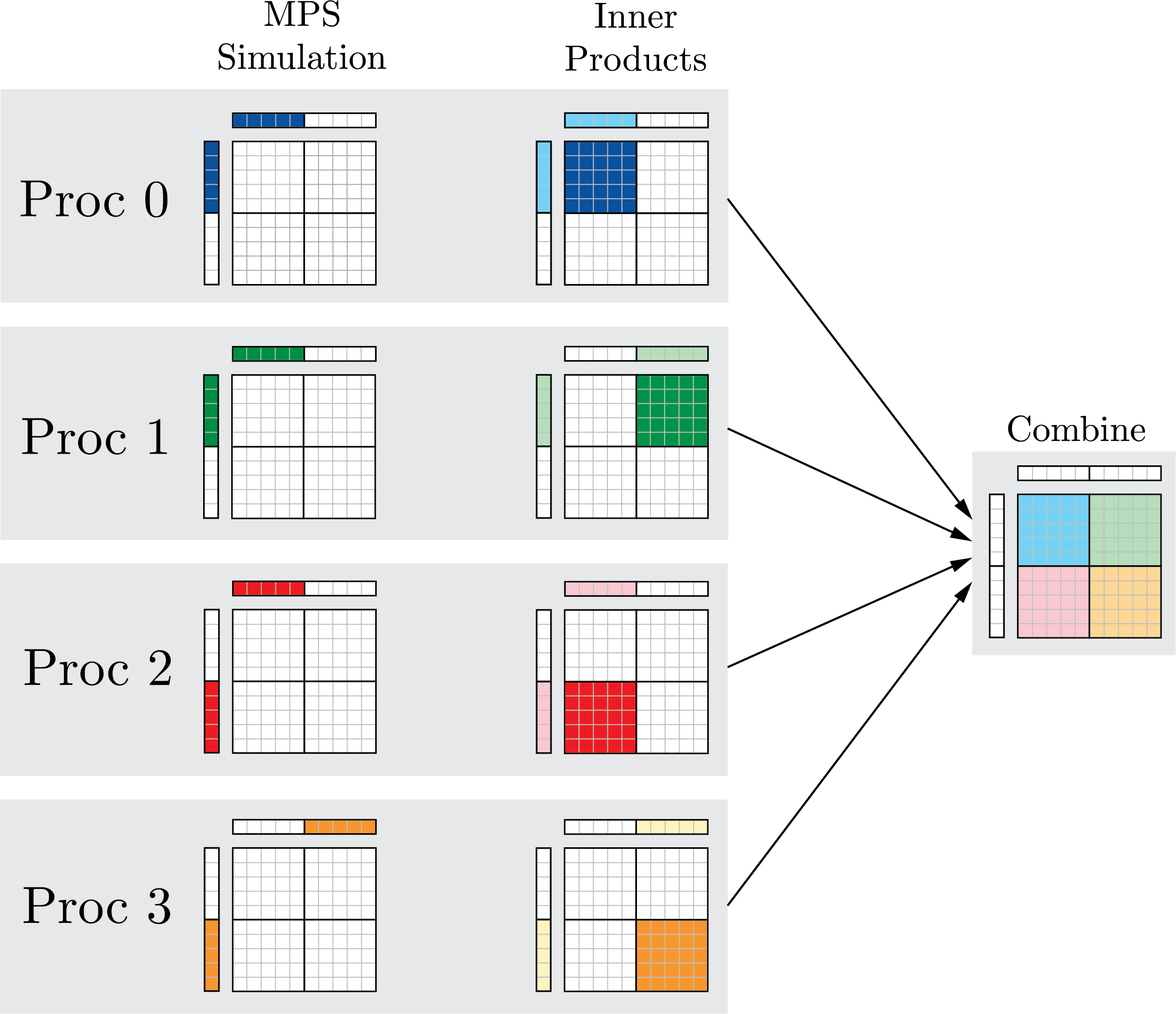}};
        \node[right=12mm of nm.east] (rr) {\includegraphics[scale=0.33]{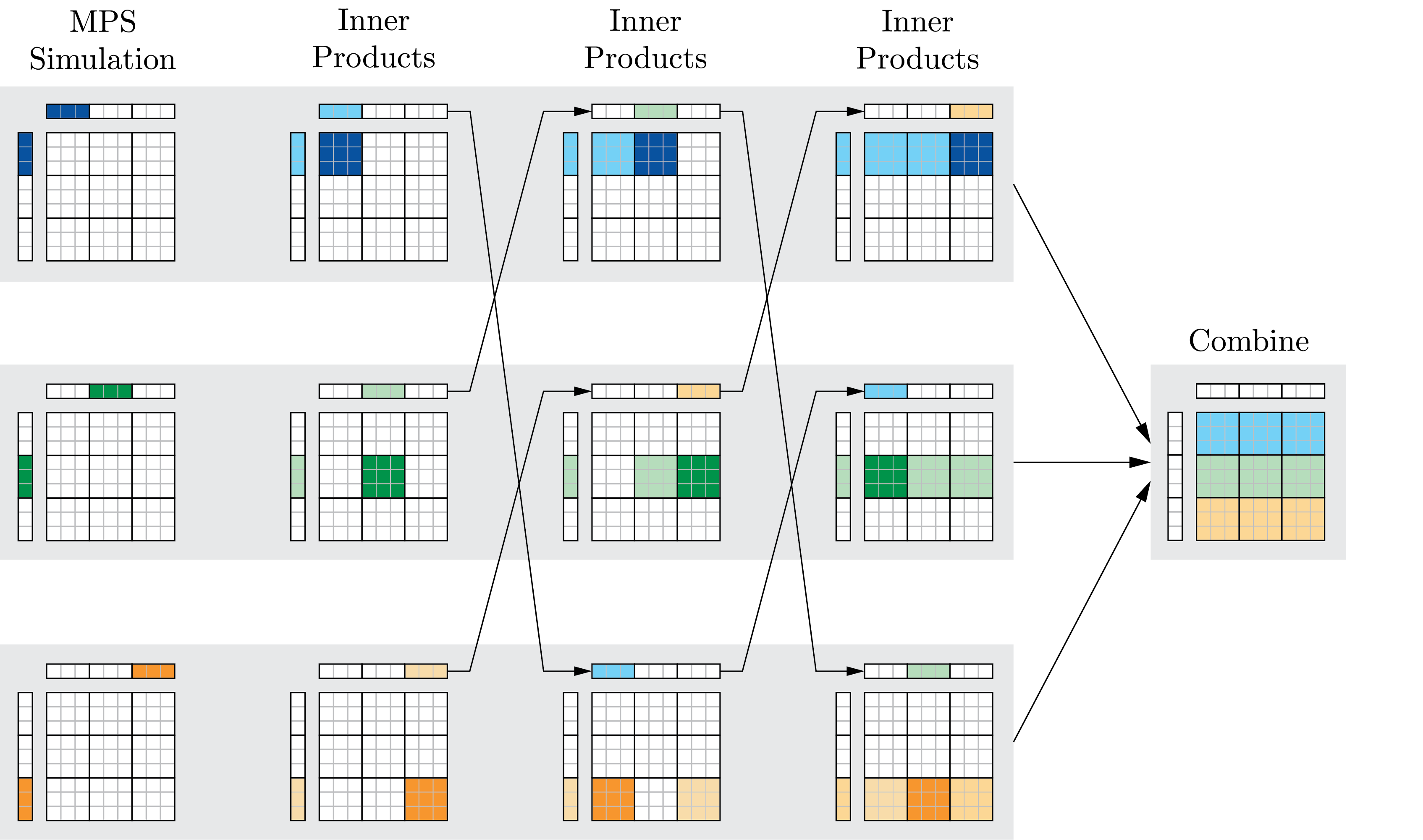}};
        \node[above left=-5mm and 0mm of nm.north west] (a) {(a)};
        \node[above left=-5mm and 0mm of rr.north west] (b) {(b)};
    \end{tikzpicture}
    \caption{Distributed computation of the Gram matrix. Two strategies are presented: (a) no-messaging  and (b) round-robin. The figure displays the sequence of steps of either strategy, along with the data held by each of the parallel processes. The square grid on each step represents the collection of entries of the Gram matrix. The vertical and horizontal arrays on each step represent the collection of quantum states required to compute the matrix entries. A white square indicates the process does not hold the corresponding data on that step. The colour a square is filled with indicates which process computed the data originally; a darker hue is used to indicate the data is being computed at the corresponding step. Arrows indicate message passing between processes.}
    \label{fig:parallel}
\end{figure*}

The computation of the Gram matrix is embarrassingly parallelizable, since each of its entries $\boldsymbol{K}_{ij} = \vert\langle \psi(\boldsymbol{x}_i),\psi( \boldsymbol{x}_j)\rangle\vert^2$ can be computed independently from each other.
As shown in Fig.~\ref{fig:parallel}, we divide the kernel matrix into tiles and issue a tile to each of the available processes.
Each process solves two tasks: first, simulate all of the circuits required to compute the entries within its tile; second, calculate the inner product of every pair of such states to obtain each of the entries.
We refer to the trivial approach depicted in Fig.~\ref{fig:parallel}a as a \textit{no-messaging} strategy since it is implemented without communication between the different processes.
It has the disadvantage that---when using $k$ processes and assuming square tiling---every single circuit will be simulated independently on $\bigO{\sqrt{k}}$ different processes, since all tiles on the same column $i$ require access to the state $\lvert\psi(\boldsymbol{x}_i)\rangle$ to compute their entries.

Alternatively, we consider the use of a \textit{round-robin} parallelization strategy, depicted in Fig.~\ref{fig:parallel}b.
First, the collection of circuits is split evenly between all of the available processes, so that each circuit is simulated only once.
Then, each process calculates all inner products between the pairs of states that resulted from local simulations.
The next step has process $i$ send half of its states to process $i-1$, which then computes all inner products between its states and the ones it just received.
This is repeated until all of the entries of the Gram matrix have been computed, when each process has calculated its own row of tiles, as shown in Fig.~\ref{fig:parallel}b.
Unlike the no-messaging strategy, each of the simulated quantum states is stored in a single process at a time, making the round-robin strategy more memory efficient.
Furthermore, round-robin is also faster if the overhead of communicating states between processes is lower than MPS simulation time (which is indeed the case, as shown in Fig.~\ref{fig:runtime_scaling}).
At the end of either strategy, every process sends the kernel entries it computed to a common process that combines them all into a single matrix.

Both distribution strategies split the computation of the Gram matrix into tiles; square tiles are favoured since they offer the best trade-off between area---number of entries---and side length---number of simulations or messages required.
We can use the fact that the Gram matrix on the training data set is symmetric to reduce the amount of entries to be computed by a half.
In contrast, the kernel matrices for inference are rectangular since there are more training data points than test data points. This makes no difference to the no-messaging strategy, but it has some subtle consequences for round-robin: in the case that there is one more row of tiles than columns, the process in charge of computing the last row will---at every step---calculate a tile in the same column as some other process and, consequently, require a copy of the subset of MPS necessary to compute said column.
Our solution to this is straightforward: if the matrix has $\ell$ columns of tiles, we arrange the $k$ available processes in groups of $\ell$ of them, and each group computes its collection of $\ell \times \ell$ tiles using the same approach as in Fig.~\ref{fig:parallel}b.
The remaining $k \mod \ell$ processes receive at each step---via an additional round of message passing---a copy of the subset of MPS they require from the matching processes within the first group.

We implemented both strategies using MPI via the \texttt{mpi4py} Python library.
Due to ITensors being a Julia package, we found some challenges orchestrating the distribution of states from the round-robin strategy at the Python level.
Ideally, we would implement the parallelisation at the Julia level as well.
However, we found the no-messaging strategy to be sufficient for small experiments, and we only implemented the round-robin strategy for the approach using \ourlibrary{} which we evaluate at larger scales (see section~\ref{sec:results_qml}).
As such, it would not be fair to compare the total runtime of ITensors against \ourlibrary{} on a distributed setting.
Instead section~\ref{sec:results_sim} compares their performance on the computation of entries on a single process.

\section{Results}

In section~\ref{sec:results_sim} we evaluate the resource scaling of our quantum kernel framework.
We study the effect of increasing the qubit interaction distance of our circuit ansatz which, as expected, increases runtime exponentially.
The scaling is different in CPUs and GPUs, and we identify a crossover point beyond which the use of GPUs is advantageous.
Both memory and runtime of MPS methods enjoy a gentle scaling on the number of qubits and, as such, the experiments reported in section~\ref{sec:results_sim} use circuits on $100$ or more qubits---more than twice the number achievable by state vector simulators~\cite{ShorSimulated}.
Section~\ref{sec:results_qml} studies the QML model performance at the scale enabled by MPS simulation.
All of the experiments in both of these sections use the circuit ansatz described in section~\ref{sec:ansatze}, instantiated with data entries $\boldsymbol{x}$ from the publicly available Elliptic data set hosted on Kaggle~\cite{elliptic}.
Experiments using the ITensors backend run on AMD EPYC 7763 CPUs, while our custom GPU backend uses NVIDIA A100 GPUs. 

\subsection{GPU advantage and resource scaling}
\label{sec:results_sim}

\begin{figure*}
    \centering
    \begin{tikzpicture}
        \node (sim) {\includegraphics[scale=0.37]{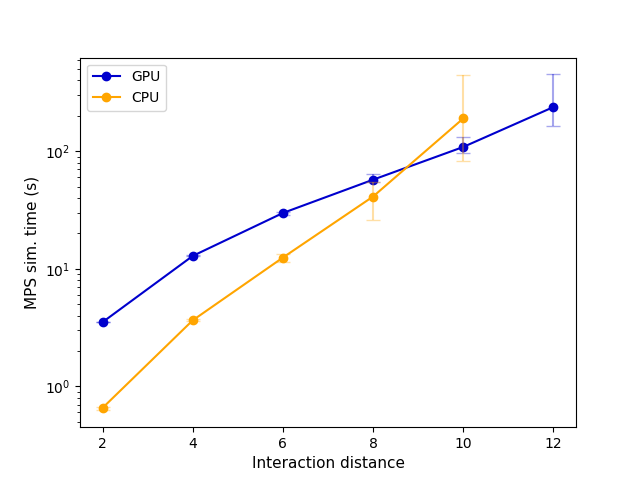}};
        \node[right=5mm of sim.east] (dot) {\includegraphics[scale=0.37]{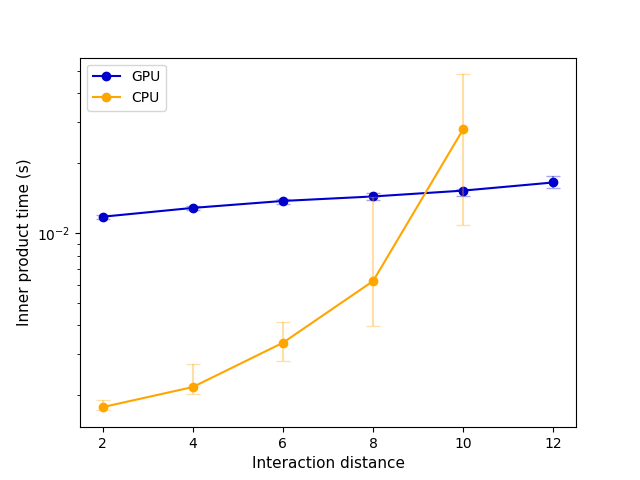}};
        \node[above left=-7mm and -7mm of sim.north west] (a) {(a)};
        \node[above left=-7mm and -7mm of dot.north west] (b) {(b)};
    \end{tikzpicture}
    \caption{Runtime scaling as the qubit interaction distance is increased. CPU corresponds to ITensors, GPU corresponds to our library. Each circuit comprises $m=100$ qubits, $r=2$ layers and uses $\gamma=1.0$. Data points shown are the median of $8$ samples in the case of MPS simulation, and of $28$ samples in the case of inner products; see the main text for details. Error bars indicate the first and third quartiles.}
    \label{fig:crossover}
\end{figure*}

By varying the value of the interaction distance $d$ in our circuit ansatz we can increase the amount of entanglement between the qubits in the final state, thus increasing the runtime and memory requirements of MPS simulation and inner product calculation (see section~\ref{sec:simulation}).
To identify the regime where GPUs provide an advantage to our QML task, we fix all circuit ansatz hyperparameters except the qubit interaction distance and take multiple samples of the time to simulate a single circuit (Fig.~\ref{fig:crossover}a) and to calculate a single inner product (Fig.~\ref{fig:crossover}b).
To obtain these samples, we execute our quantum kernel framework once for each value of the interaction distance, obtaining in each instance a Gram matrix~\eqref{eq:Gram} on a subset of size $8$ of the Elliptic Bitcoin data set---this corresponds to the simulation of $8$ different circuits and the calculation of $28$ inner products per choice of interaction distance.
Fig.~\ref{fig:crossover} displays the median and quartiles of these samples,\footnote{The first time ITensors' code is called from Python there is a large overhead due to precompilation of Julia code. Consequently, we take the median of the samples instead of the mean, since the latter would have biased the results against ITensors due to this one-time overhead. All samples are taken on the same run, so only the first few of them are affected by this overhead. Runtime of the ITensors library is measured within Julia.} both for ITensors and for \ourlibrary{}.
In both tasks, there is a crossover point between $d=8$ and $d=10$ beyond which our GPU implementation becomes faster than ITensors running on CPUs.
For smaller interaction distances the CPU backend is favoured; this is likely caused by overheads in our GPU backend, either due to implementation details in our library (e.g. written in Python) or more fundamental CPU-GPU communication bottlenecks.

Remarkably, Fig.~\ref{fig:crossover}b shows a dramatic difference in runtime between the two backends for the task of calculating inner products.
Since the calculation of inner products becomes the bottleneck of our quantum kernel framework for large enough data sets (see Fig.~\ref{fig:runtime_scaling}), the use of GPUs is crucial when using an interaction distance of $d \geq 10$.
However, as discussed in section~\ref{sec:results_qml}, our results are inconclusive on whether the model's classification performance benefits from a larger interaction distance.
Consequently, we recommend users of our framework to carefully analyze whether their circuit ansatz lies within the CPU-favoured or GPU-favoured regime.
An effective way to inform such a decision is to observe the virtual bond dimension $\chi$ of the MPS at the end of the simulation, since this is the main contributor to the runtime complexity $\bigO{m\chi^3}$ of both inner product calculation and MPS simulation ($m$ being the number of qubits).
Table~\ref{tab:crossover_chi} displays the (average of) largest virtual bond dimension $\chi$ of the final MPS for each of the data points in Fig.~\ref{fig:crossover}; from it we conclude that $\chi \geq 320$ is enough to enter the GPU-favoured regime.

Table~\ref{tab:crossover_chi} displays $\chi$ for both the CPU and the GPU backend.
Since both backends use the same MPS simulation algorithm, we should expect their bond dimensions match.
Indeed, this is the case---up to a negligible variation\footnote{This may be due to the aggregation of subtle optimisations done in our library that are not present in ITensors. An example of this is the SVD decomposition of two-qubit gates \textit{before} their application which, in the case of $\Rxx$ gates, causes its two $0$ singular values to be truncated in advance. Otherwise, half of the singular values in the final step of Fig.~\ref{fig:gate_application}b should be $0$, but the existence of floating point errors causes these not to be exactly $0$, which may lead to some of these `essentially zero' singular values not being truncated.}---which corroborates that the difference in runtime displayed in Fig.~\ref{fig:crossover} is caused solely by the implementation details and computing architecture of either backend.
Table~\ref{tab:crossover_chi} also displays the average memory footprint of each MPS, as reported by our GPU backend---this is roughly the same for the CPU backend, differing only due to the slightly larger virtual bond dimension.

\begin{table} 
\caption{Average of the largest bond dimension for points in Fig.~\ref{fig:crossover}.}
\label{tab:crossover_chi}
\begin{center}
\begin{tabular}{|c|r|r|c|} 
 \hline
 \textbf{interaction} & \textbf{Avg. largest} & \textbf{Avg. largest} & \textbf{Memory per} \\
 \textbf{distance} & \textbf{$\chi$ (GPU)} & \textbf{$\chi$ (CPU)} & \textbf{MPS (MiB)} \\
 \hline
 2 & 10.125 & 10.250 & 0.02 \\ 
 4 & 28.625 & 29.375 & 0.15 \\ 
 6 & 71.875 & 73.625 & 1.07 \\
 8 & 137.125 & 137.125 & 4.39 \\
 10 & 320.125 & 326.750 & 20.12 \\
 12 & 595.625 & --- & 106.35 \\
 \hline
\end{tabular}
\end{center}
\end{table}

\begin{figure}
    \centering
    \includegraphics[scale=0.55]{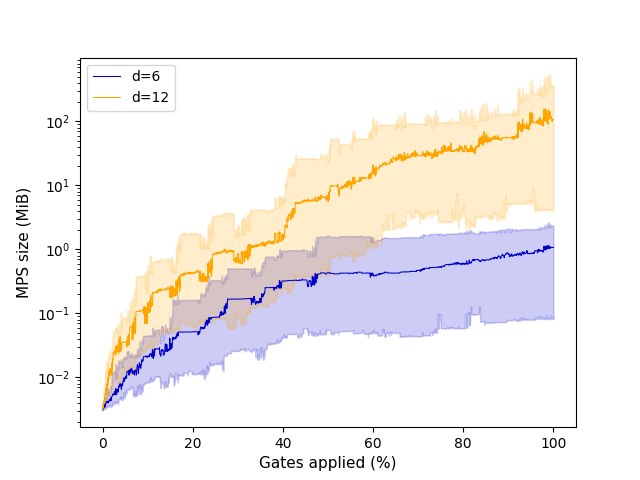}
    \caption{Memory required to store the MPS throughout the simulation of a circuit. The x-axis indicates the progress of the simulation as the percentage of gates already applied. Two families of circuits are considered: one with interaction distance $d=6$ (blue) and another with $d=12$ (orange); both with $m=100$ qubits, $r=2$ layers and $\gamma=1.0$. Each thick line corresponds to the mean of $8$ samples from the corresponding circuit family; the samples are the same ones used to generate the data points at $d=6$ and $d=12$ in Fig.~\ref{fig:crossover}. The shaded areas are bounded from the bottom (top) by the sample that required the least (most) memory.}
    \label{fig:mem_evolution}
\end{figure}

As shown in Fig.~\ref{fig:crossover}, the runtime of MPS simulation scales exponentially with respect to the qubit interaction distance.
This is to be expected, since the bond dimension increases exponentially with the number of gates applied, and for larger interaction distances we are applying more $\Rxx$ gates (see Fig.~\ref{fig:ansatze}) and we require more $\SWAP$ gates.
Fig.~\ref{fig:mem_evolution} plots the evolution of the average memory required to store the MPS throughout simulation, for two circuits with different interaction distance.
The sharp drops in bond dimension in Fig.~\ref{fig:mem_evolution} are due to SVD truncation; without it, the memory required by our framework would quickly become unmanageable, more so considering that for larger scale applications (see Fig.~\ref{fig:runtime_scaling}) we store thousands of different MPS in memory to compute each of the entries of the Gram matrix~\eqref{eq:Gram}.
Recall that we bound the truncation error~\eqref{eq:trunc_error} below $10^{-16}$ and, hence, the inaccuracy due to approximation is negligible.
Nonetheless, Fig.~\ref{fig:mem_evolution} shows an exponential increase in memory with respect to the number of gates applied, which is characteristic of tensor network methods~\cite{MPS_sim, Tindall2024, Gray2021}.

\begin{figure}
    \centering
    \includegraphics[scale=0.55]{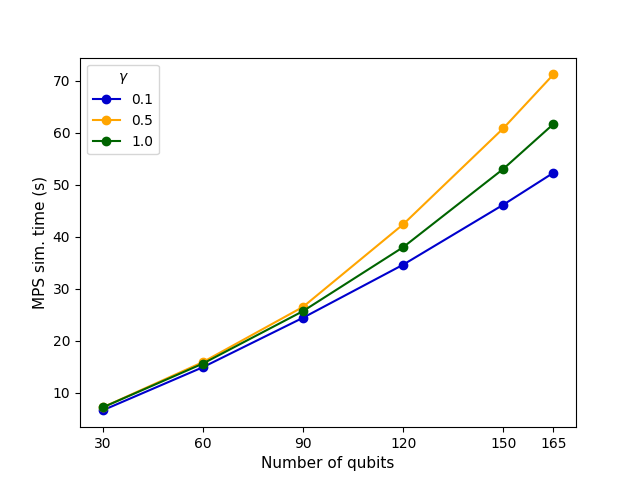}
    \caption{Simulation time for circuits with varying number of qubits. Circuits contain $r=2$ layers and use interaction distance $d=6$. Each point in the plot corresponds to the average of $8$ samples initialised from randomly chosen rows $\boldsymbol{x}$ from the Elliptic Bitcoin data set. The scaling for different values of $\gamma$ is shown.}
    \label{fig:qubit_scaling}
\end{figure}

Fig.~\ref{fig:qubit_scaling} illustrates the runtime scaling of MPS simulation with respect to the number of qubits.
The asymptotic scaling of MPS simulation is $\bigO{m\chi^3}$ where $m$ is the number of qubits, so it would be natural to expect a linear correlation in Fig.~\ref{fig:qubit_scaling}.
In practice, however, we find that the bond dimension $\chi$ is not independent from $m$.
Nevertheless, the scaling remains manageable, allowing us to simulate our circuit ansatz with up to $165$ qubits with ease.
Fig.~\ref{fig:qubit_scaling} also shows how the hyperparameter $\gamma$ affects the runtime performance of our framework.
Intuitively, we can expect the strength of the entanglement generated by the circuit to be dependent on the value of $\gamma$, since it is a factor in the angle of the $\Rz$ and $\Rxx$ gates; see equations~\ref{eq:Hz} and~\eqref{eq:Hxx}.
Notice that the largest runtime corresponds to the intermediate value of $\gamma=0.5$: the other two values of $\gamma$ often lead to $\Rxx$ gates with angles closer to $0$ and $\pi$---which correspond to Pauli gates---and, hence, generate weaker entanglement.
Similar plots to those of Fig.~\ref{fig:qubit_scaling} are obtained when displaying the runtime of inner product calculation in the y-axis instead of simulation time; this is to be expected since both have the same runtime complexity of $\bigO{m\chi^3}$.

\begin{figure}
    \centering
    \includegraphics[scale=0.55]{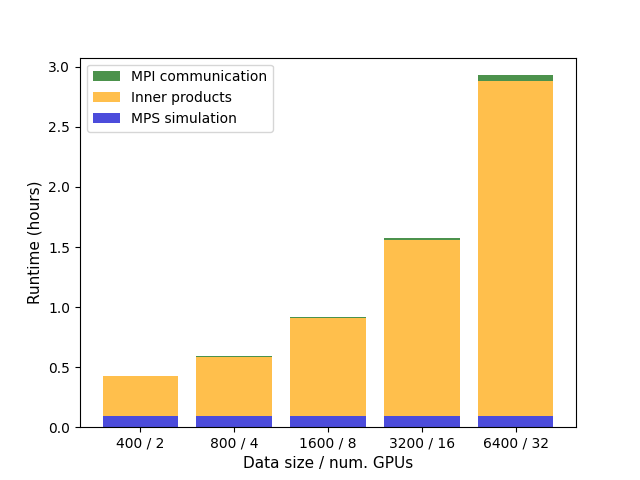}
    \caption{Breakdown of wall-clock runtime to compute the Gram matrix for the training data set. Each bar corresponds to a data set twice as large as the previous bar, using twice as many parallel GPUs. We use our GPU simulation library along with the round-robin parallelization strategy. Circuit ansatz uses $165$ qubits (features), $r=2$ layers, interaction distance $d=1$ and $\gamma=0.1$. }
    \label{fig:runtime_scaling}
\end{figure}

We now explore the performance of our framework with 165 qubits as we increase the number of parallel processing units.
Fig.~\ref{fig:runtime_scaling} illustrates the embarrasingly parallel nature of the task at hand: increasing both the number of data points in the training data set and the number of processing units available by the same factor does not change the amount of time spent in simulation when using the round-robin strategy described in section~\ref{sec:parallel}.
Recall that the number of inner products calculated to produce the Gram matrix~\eqref{eq:Gram} on the training data set scales quadratically with the number of entries; as such---unlike in the case of simulation---a linear increase in the number of parallel processing units is not enough to maintain the wall-clock runtime constant.
Indeed, we observe the expected increase in runtime in Fig.~\ref{fig:runtime_scaling}: for twice as many entries in the data set, the total runtime is increased by roughly\footnote{Since the Gram matrix is symmetric, $N$ entries require $N(N-1)/2$ inner products to be calculated, rather than $N^2$. This is what causes the inner product bars in Fig.~\ref{fig:runtime_scaling} to increase by slightly less than a factor of two.} a factor of four which, when distributed across twice as many parallel processors, equates to each of them taking twice as long.
These results can be extrapolated to a larger scale where the training on a data set of $64,000$ entries could be achieved in $30$ hours using $320$ GPUs, or in $15$ hours using $640$ GPUs.

The circuit ansatz used to obtain the data in Fig.~\ref{fig:runtime_scaling} coincides with that of Fig.~\ref{fig:AUC_train} and Fig.~\ref{fig:AUC_test}, discussed in the next section.
It uses an interaction distance of $d=1$ and, hence, using the CPU backend with ITensors would be favourable, as suggested by Fig.~\ref{fig:crossover}.
Each MPS on $165$ qubits for this circuit ansatz requires less than $15$ KiB of memory on average.
Consequently, memory is unlikely to be a bottleneck for scalability when using this particular circuit ansatz, since storing all of the quantum states for a $64,000$ data set would require less than $1$ GiB.
However, it is crucial to remark that the memory required by an MPS scales as $\bigO{m\chi^2}$ where $m$ is the number of qubits and $\chi$ is an upper bound of the virtual bond dimensions.
In the case of the circuit ansatz used in Fig.~\ref{fig:runtime_scaling}, $\chi$ is approximately $2$, which explains its extremely low memory requirements and relatively fast computation.
In comparison, as reported in Fig.~\ref{fig:crossover}, larger interaction distances will rapidly increase the runtime and memory requirements due to a much larger value of $\chi$ (see Table~\ref{tab:crossover_chi}).
Therefore, the usage of more complex circuit ansatze must be justified by a noticeable increase in classification performance; we explore this topic in the next section (see Table~\ref{tab:svm_diff_topologies}).

Once the Gram matrix is constructed for the training data set, classification of a single unlabeled data point requires us to simulate the corresponding new circuit, calculate the inner products of the resulting state with each of the states from the training data set and feed these along with the Gram matrix to a standard SVM pipeline.
Assuming the MPS of each of the quantum states from the training stage are stored in memory across different processors, the calculation of inner products can be achieved in parallel and will scale linearly with the training data set size.
In the case of the circuit ansatz used in Fig.~\ref{fig:runtime_scaling}, each inner product requires approximately $0.02$ seconds, which when scaled to an application with $64,000$ training size and $320$ GPUs would equate to $4$ seconds.
MPS simulation for the corresponding new data point using this circuit ansatz requires an additional $2$ seconds and does not benefit from parallelization in the current framework.
The runtime may be lowered for this particular circuit ansatz using the ITensors backend on CPUs, but for more complex ansatze the total time would increase---as shown in Fig.~\ref{fig:crossover}---and the use our GPU backend would become advantageous.

\subsection{Quantum Machine Learning Performance}
\label{sec:results_qml}

\begin{figure}
    \centering
    \includegraphics[width=0.5\textwidth]{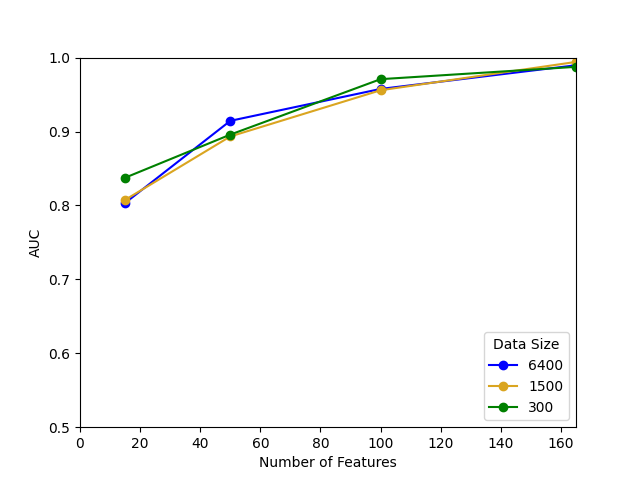}
    \caption{Training set Area Under the Curve prediction score with 6400 (blue), 1500 (gold), and 300 (green)  data sample sizes on the Elliptic Bitcoin data set with increasing feature number.}
    \label{fig:AUC_train}
\end{figure}
\begin{figure}
    \centering
    \includegraphics[width=0.5\textwidth]{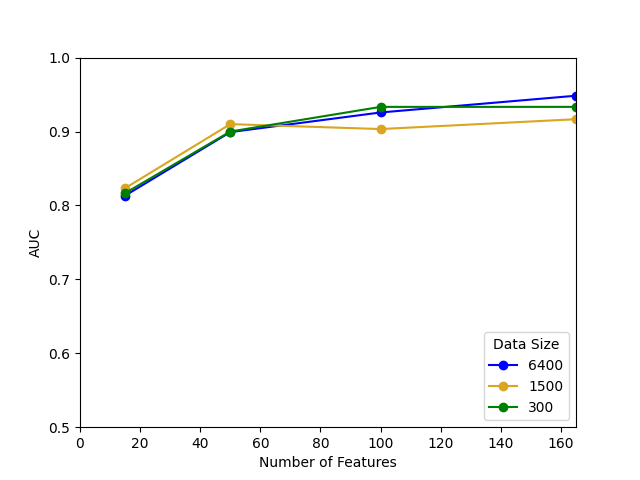}
    \caption{Test set Area Under the Curve prediction score with 6400 (blue), 1500 (gold), and 300 (green)  data sample sizes. on the Elliptic Bitcoin data set with increasing feature number.}
    \label{fig:AUC_test}
\end{figure}

We model performance with a quantum kernel computed using our simulator, with the SVM model applied to the Elliptic Bitcoin data set. This data set has 165 features corresponding with 4,545 data points labeled illicit and 42,019 data points labeled licit transactions, allowing us to push our simulator to regimes modern quantum computers have yet to reach, and well beyond what state vector simulators can support.  We aim to explore classification capabilities at data volumes and feature dimensions unlocked through our simulator framework. We further aim to assess how increasing expressivity through the interaction distance or number of circuit layers affects the model's classification performance. Data is prepared using a standard data engineering pipeline to normalize and scale the data. Data samples are down selected and seeded to a specified dimension with balanced data and an 80/20 train-test split. SVM regularization parameter $C \in [0.01, 4]$ and the tolerance for SVM is $10^{-3}$. Model performance is assessed using the standard metrics including accuracy, recall, precision, and Area Under the Curve (AUC). The AUC metric quantifies the area under the Receiver Operating Characteristic (ROC) curve where the y-axis has the true positive rate and x-axis has the false positive rate~\cite{ROC}. AUC can capture a classifier's capability to distinguish between true positives and false positives at a given threshold, so this metric is primarily used to quantify classification performance.

We assess SVM performance with a quantum kernel at a previously unachieved scale, which provides insight into the interplay of large feature dimension and training data size in achieving quality results on the test data set.  The train data kernel and the test data kernel are supplied to the standard SVM classifier for data sample sizes of 300, 1500, and 6400 with 15, 50, 100, and 165 features. We use a linear ansatz with interaction distance of one ($d=1$), two encoding circuit repetitions ($r=2$), and $\gamma = 0.1$. The quality of training is displayed in Fig.~\ref{fig:AUC_train} by testing how well the trained SVM predicts the correct labels of the training data set. More importantly, we assess how the trained SVM correctly predicts on the test data set in Fig.~\ref{fig:AUC_test}. It is clear both the training and test prediction quality continues to improve as more features are added. The prediction score on the training data set achieves its highest value when trained on 300 samples, which is likely an indicator of model overfitting. This is observed in the prediction score of the test data set: the test score for the data sample size of 300 does not improve consistently with the addition of more features with an equivalent 0.933 AUC score with both 100 features and 165 features. In contrast, we see consistent classification quality improvement on the test data set with data sample size 6400 with a $2.44\%$ AUC improvement using 165 features over 100 features. This observation is indicative that more training data is needed to avoid model overfitting as the feature dimension increases. Scaling to larger amounts of training data is achievable with our parallelization strategy and the addition of more processors; see Fig.~\ref{fig:runtime_scaling}.  
These large-scale simulations reveal that model performance is improved with the addition of more qubits (i.e. more features) and the addition of more training data which is a notable and promising result for near term quantum machine learning algorithms. While simulation scale is feasible, careful consideration is needed to see how quantum circuit complexity affects model performance.

\begin{table} 
\caption{SVM performance with 50 features and 2 encoding circuit repetitions. Experiment with highest AUC marked in bold.}
\label{tab:svm_diff_topologies}
\begin{center}
\begin{tabular}{ |c|c|c|c|c|c|c| } 
 \hline
 \textbf{kernel} & \textbf{d} & \textbf{$\gamma$} & \textbf{AUC} & \textbf{Recall} & \textbf{Precision} &\textbf{Accuracy} \\ 
\hline
  Gaussian & - & - & 0.892 & 0.883 & 0.898 & 0.892 \\
\hline
  quantum & 1 & 0.1 & 0.877 & 0.883 & 0.874 & 0.877 \\
  quantum & 2 & 0.1 & 0.877 & 0.883 & 0.874 & 0.877 \\
  quantum & 4 & 0.1 & 0.877 & 0.883 & 0.874 & 0.877 \\
  quantum & 6 & 0.1 & 0.877 & 0.883 & 0.874 & 0.877 \\
  quantum & 1 & 0.5 & 0.902 & 0.933 & 0.879 & 0.902 \\
  quantum & 2 & 0.5 & 0.900 & 0.946 & 0.867 & 0.900 \\
  quantum & \textbf{4} & \textbf{0.5} & \textbf{0.904} & \textbf{0.946} & \textbf{0.874} & \textbf{0.904} \\
  quantum & 6 & 0.5 & 0.881 & 0.937 & 0.843 & 0.881 \\
  quantum & 1 & 1.0 & 0.898 & 0.950 & 0.863 & 0.898 \\
  quantum & 2 & 1.0 & 0.894 & 0.942 & 0.861 & 0.894 \\
  quantum & 4 & 1.0 & 0.898 & 0.962 & 0.853 & 0.898 \\
  quantum & 6 & 1.0 & 0.887 & 0.975 & 0.831 & 0.887 \\
\hline
\end{tabular}
\end{center}
\end{table}

Expressivity is quantified by circuits that require a higher bond dimension, which is induced by longer qubit interaction distances, more circuit layers, and the value of the $\gamma$ coefficient. In section~\ref{sec:results_sim} we illustrate how tuning these circuit ansatz hyperparameters affects the resource requirements of our framework. We now report on experiments aimed to elucidate the effect of circuit complexity on model performance using $50$ features. Each run consists of 6 data samples and the metrics are averaged over the 6 runs with the same regularization parameter. Each implementation with different specifications is independent of previous runs, and the best performing model for a given regularization coefficient is chosen independently. Our goal is to determine which attribute -- interaction distance, depth, or $\gamma$ -- has the largest effect on model performance. We experiment with the circuit ansatz for $d \in (1,6)$ and using coefficients $\gamma = (0.1,0.5,1)$.  Machine learning performance metrics are compared to the SVM algorithm with the common Gaussian kernel,
\begin{equation}
 e^{-\alpha |\textbf{x} - \textbf{x}^\prime|^2}.
\end{equation}
We choose the bandwidth parameter such that $\alpha = 1/[m\cdot\text{var(X)}]$ for some data set $X$. The regularization parameter range and tolerance for SVM with a Gaussian kernel are equivalent to the quantum kernel experiments. Performance on the test data set using these models is detailed in Table~\ref{tab:svm_diff_topologies}. None of the models with $\gamma = 0.1$ outperformed the classical model. Further increasing interaction distance for small $\gamma$ has little impact on model performance which is likely due to very small interaction coefficients in the encoding Hamiltonian. The quantum model did outperform the classical model with $\gamma \in \{0.5,1\}$, and the interaction distance impact on model quality is more clear as the larger two qubit interaction coefficients generate more correlation. We see $d=4$  demonstrates the best performance with $\gamma = 0.5$ while $d=1$ demonstrates the best performance with $\gamma = 1$. Model performance diminishes when $d=6$ and $\gamma \in \{0.5,1\}$ which results from an overly expressive model that overfits the train data set. The effect of kernel bandwidth ($\gamma$) and the inclusion of more generators ($d$) on model performance indicates there is a parameter regime to achieve better quality generalization for the quantum model which will likely have a data set dependence. 

\begin{table} 
\caption{Ansatz repetition effect on SVM performance on 50 features $d=1$ and $\gamma = 1$. Experiment with highest AUC marked in bold.}
\label{tab:svm_diff_depths}
\begin{center}
\begin{tabular}{ |c|c|c|c|c| } 
 \hline
 \textbf{depth}  & \textbf{AUC} & \textbf{Recall} & \textbf{Precision} &\textbf{Accuracy} \\ 
 \hline
  2 & 0.898 & 0.950 & 0.863 & 0.898 \\
  \textbf{4} & \textbf{0.900} & \textbf{0.975} & \textbf{0.849} & \textbf{0.900} \\
  8 & 0.844 & 0.996 & 0.765 & 0.844 \\
  12 & 0.810 & 1.000 & 0.727 & 0.810 \\
  16 & 0.810 & 0.987 & 0.732 & 0.810 \\
  20 & 0.798 & 0.992 & 0.717 & 0.798 \\
 \hline
\end{tabular}
\end{center}
\end{table}

Circuit depth, induced by repeated ansatz repetitions, further increases the simulation complexity and model expressivity. We assess how circuit depth influences both train and test performance for the best performing configuration above. Increasing the depth leads to kernel concentration, which is known to cause model untrainability~\cite{ExpConc}; this phenomenon is apparent in the results shown in Table~\ref{tab:svm_diff_depths}. Increasing the depth significantly diminishes model performance on the test data set. Intuitively, more applications of the ansatz rotates data points farther away from each other, such that the overlaps become increasingly small; therefore, no useful information is extracted from the feature map. 

\section{Conclusion}

We present a tensor network simulation framework designed to scale machine learning applications using a quantum kernel. Two simulation architectures are presented, one using the ITensors library on CPUs and the second using our own tensor network library, \ourlibrary{}~\cite{pytket-cutensornet}, built on top of NVIDIA cuTensorNet for GPUs. Performance of these simulators is assessed for the two subroutines needed for quantum kernel applications: MPS simulation of quantum circuits and calculation of the overlap of two MPS. Considering a linear chain qubit connectivity where each qubit interacts with its nearest $d$ neighbors, we find a crossover point when $d=10$ and $\chi=320$ after which the GPU simulator outperforms the CPU simulator. We further articulate compute requirements to train a classifier with $>100$ features and $>10,000$ training data points as well as inference on unlabeled data points. These results showcase that quantum kernel methods are implementable at scale with reasonable compute times using enough CPUs or GPUs. 

The scalability limits of our method were addressed throughout the paper. Explicitly, Table~\ref{tab:crossover_chi} and Fig.~\ref{fig:mem_evolution} show how memory requirement scales with circuit ansatz complexity, Fig.~\ref{fig:qubit_scaling} shows the runtime scaling with respect to number of qubits (features), and Fig.~\ref{fig:runtime_scaling} shows the runtime scaling as we increase the training data size (and the number of parallel processors). We find our quantum kernel method outperformed the baseline Gaussian kernel in Table~\ref{tab:svm_diff_topologies}demonstrating model performance alongside implementation performance for machine learning tasks. Ultimately, the scalability limits are imposed by the quadratic increase in number of kernel matrix entries with respect to data size. The only exponential scaling in our method comes from simulation of increasingly complex circuit ansatz (see Fig.~\ref{fig:crossover}), but in section~\ref{sec:results_qml} we show evidence that a simple circuit ansatz (and hence, cheap simulation) is sufficient to obtain good classification performance. We reiterate that our simulations are (virtually) noiseless: as discussed immediately after equation~\eqref{eq:trunc_error}, we configure truncation to only remove singular values below $10^{-16}$ and, consequently, errors due to truncation are at the same order of magnitude as floating point errors for 64-bit precision. If future work shows that using more complex circuit ansatze is beneficial, more aggressive truncation may be deemed necessary for scalability purposes. In such a situation, analysis of the noise induced by truncation would be necessary.

The simulator is used to perform one of the largest quantum kernel implementations to date. These large scale simulations show that the quantum model yields better test scores with the inclusion of both more features and more training data. The CPU-GPU crossover point illustrates GPU advantage in tensor network simulation for circuits of a given complexity, and we assess how this complexity, which translates to quantum feature map expressivity, affects classification performance. Our numerical experiments show more expressive circuits cause model overfitting; therefore the CPU architecture is most efficient for simulating the quantum kernel method with the Elliptic data set. Other data sets (e.g. quantum data) might benefit from more complicated interaction graphs necessitating the GPU based simulator. Regardless, our quantum kernel simulation engine enables researchers and practitioners to implement quantum models at industry relevant scales making quantum machine learning applications practical in the near-term.

\section*{acknowledgment}

The authors thank Matthias Rosenkranz and Iakov Polyak for useful discussions and engineering support.

This research used resources of the National Energy Research Scientific Computing Center (NERSC), a Department of Energy Office of Science User Facility using NERSC award DDR-ERCAP0029782.

DISCLAIMER: This paper was prepared for information purposes and is not a product of HSBC Bank Plc. or its affiliates. Neither HSBC Bank Plc. nor any of its affiliates make any explicit or implied representation or warranty, and none of them accept any liability in connection with this paper, including, but not limited to, the completeness, accuracy, reliability of information contained herein and the potential legal, compliance, tax, or accounting effects thereof. This document is not intended as investment research or investment advice, or a recommendation, offer or solicitation for the purchase or sale of any security, financial instrument, financial product or service, or to be used in any way for evaluating the merits of participating in any transaction.

\bibliography{IEEEabrv, references}

\clearpage
\newpage

\appendix[Artifact Descripion]

\section{Overview of Contributions and Artifacts}

\subsection{Paper's Main Contributions}

The contributions of our paper can be classified in two groups.
The first group of contributions ($C_1$) refers to the development and benchmarking of a framework for supervised machine learning based on a Quantum Kernel method, employing MPS simulators to run the quantum computations on GPUs and CPUs. 
We list its main characteristics as separate contributions.

\begin{description}
    \item[$C_{1.1}$] The use of MPS simulators allows us to scale to larger number of features (qubits) than standard statevector simulators.
    \item[$C_{1.2}$] We identify a crossover point beyond which running our framework on GPUs is advantageous over CPUs.
    \item[$C_{1.3}$] Our parallelization strategy allows us to push the scale of the machine learning task to $6400$ datapoints. Timing and resource requirements for larger datasets are easy to predict.
\end{description}

The second group of contributions ($C_2$) refers to the study of the classification performance of the machine learning models trained using our framework.
We list our main results below.

\begin{description}
    \item[$C_{2.1}$] Classification results improve with the number of features and data size, providing evidence for utility of quantum kernel methods at scale.
    \item[$C_{2.2}$] Our quantum kernel method performs better than a standard Gaussian kernel across multiple classification performance metrics.
    \item[$C_{2.3}$] Increasing the complexity of the quantum circuit ansatz (model expressivity) does not translate to an improvement in classification performance in our studied dataset.
\end{description}

\subsection{Computational Artifacts}

All artifacts and source code are archived under a single DOI: \url{10.5281/zenodo.12568631}.

The source code of the framework presented in our paper is publicly available at \url{https://github.com/PabloAndresCQ/qml-cutensornet}.
A README file is included in the root directory, providing step by step installation instructions.
Below we list the directories within this repository that include the instructions and supporting scripts to reproduce each of the figures and tables in the paper. 

\begin{description}
    \item[$A_1$] \href{https://github.com/PabloAndresCQ/qml-cutensornet/tree/main/runs/qubit_scaling}{runs/qubit\_scaling}
    \item[$A_2$] \href{https://github.com/PabloAndresCQ/qml-cutensornet/tree/main/runs/mem_evol}{runs/mem\_evol}
    \item[$A_3$] \href{https://github.com/PabloAndresCQ/qml-cutensornet/tree/main/runs/crossover}{runs/crossover}
    \item[$A_4$] \href{https://github.com/PabloAndresCQ/qml-cutensornet/tree/main/runs/runtime_scaling}{runs/runtime\_scaling}
    \item[$A_5$] \href{https://github.com/PabloAndresCQ/qml-cutensornet/tree/main/runs/qml_figures}{runs/qml\_figures}
    \item[$A_6$] \href{https://github.com/PabloAndresCQ/qml-cutensornet/tree/main/runs/table2}{runs/table2}
    \item[$A_7$] \href{https://github.com/PabloAndresCQ/qml-cutensornet/tree/main/runs/table3}{runs/table3}
\end{description}

The table below indicates the relation of computational artifacts to contributions and 
points to the elements in the paper that are reproducible by each artifact.

\begin{center}
\begin{tabular}{rll}
\toprule
Artifact ID  &  Contributions &  Related \\
             &  Supported     &  Paper Elements \\
\midrule
$A_1$   & $C_{1.1}$ & Figure 7 \\
\midrule
$A_2$   & $C_{1.1}$ & Figure 6 \\
\midrule
$A_3$   & $C_{1.2}$ & Figure 5 \\
        &           & Table 1 \\
\midrule
$A_4$   & $C_{1.3}$ & Figure 8 \\
\midrule
$A_5$   & $C_{2.1}$ & Figures 9-10 \\
\midrule
$A_6$   & $C_{2.2}$ & Table 2 \\ 
        & $C_{2.3}$ & \\
\midrule
$A_7$   & $C_{2.3}$ & Table 3 \\
\bottomrule
\end{tabular}
\end{center}

\newpage

\section{Artifact Identification}

\newartifact 

\artrel

Artifact $A_1$ generates the data for Figure 7 and plots it.
Figure 7 shows the scaling of runtime of the simulation primitive in our framework as the number of features (qubits) is increased.
It is shown that we can comfortably reach $165$ qubits, which is beyond the regime attainable by statevector simulators ($>40$), as discussed in the paper's body.
As such, it directly supports the claim of contribution $C_{1.1}$.

\artexp

The runtime required to simulate the circuit ansatz increases with the number of qubits, but it does so in a manageable way: shallow circuits on $165$ qubits are shown to be simulated in $70$ seconds or fewer.
The fact that we can simulate circuits beyond $40$ qubits is the main aspect supporting contribution $C_{1.1}$.

\arttime

Below 90 minutes to obtain all data for this artifact, using an NVIDIA A100 GPU.

\artin

\artinpart{Hardware}

Data obtained from experiments on a single NVIDIA A100 GPU from \href{https://docs.nersc.gov/systems/perlmutter/architecture/}{Perlmutter}. Any NVIDIA GPU with Compute Capability +7.0 or high-end CPU should reproduce the same qualitative scaling behaviour. Only a single device is required.

\artinpart{Software}

The source code of the framework presented in our paper is publicly available at \url{https://github.com/PabloAndresCQ/qml-cutensornet}.
Installation instructions, including version dependencies are included in the README file at the root of this repository.
\begin{itemize}
    \item General requirements: Python 3.10,  pytket 1.26.0 (\href{https://tket.quantinuum.com/api-docs/getting_started.html}{url}), pandas 2.2.1 (\href{https://pandas.pydata.org/}{url}), scikit-learn 1.4.1.post1 (\href{https://scikit-learn.org/stable/}{url}), mpi4py 3.1.5 (\href{https://mpi4py.readthedocs.io/en/stable/}{url}).
    \item The GPU backend uses cuQuantum 3.10 (\href{https://docs.nvidia.com/cuda/cuquantum/latest/getting_started/getting_started.html}{url}) and pytket-cutensornet 0.6.0 (\href{https://github.com/CQCL/pytket-cutensornet}{url}).
    \item The CPU backend uses Julia 1.9.4 (\href{https://julialang.org/}{url}), pyJulia 0.6.2 (\href{https://github.com/JuliaPy/pyjulia}{url}) and ITensors 0.3.37 (\href{https://github.com/ITensor/ITensors.jl}{url}).
\end{itemize}

\artinpart{Datasets / Inputs}

Experiments use the Elliptic Bitcoin Dataset, which needs to be downloaded from Kaggle at \url{https://www.kaggle.com/datasets/ellipticco/elliptic-data-set}. Then, run the following command to preprocess the dataset: \texttt{python elliptic\_preproc.py}.

\artinpart{Installation and Deployment}

No code compilation is required. Setup is described in detail in the README file at the root of our repository \url{https://github.com/PabloAndresCQ/qml-cutensornet}. Step by step instructions to reproduce this artifact are provided in the following section.

\artcomp

The workflow is separated in three tasks to be executed in sequence: $T_1 \rightarrow T_2 \rightarrow T_3$.

\begin{description}
    \item[$T_1$] Execute the experiments and produce the raw data. Do so by running \texttt{run\_all.sh} from the directory of this artifact (see \textit{Computational Artifacts} section).
    \item[$T_2$] Combine the output files that were produced by $T_1$ and generate a \texttt{results.csv}. Do so by running \texttt{python to\_csv.py}.
    \item[$T_3$] Produce the plot from the content of the CSV file generated by $T_2$. Do so by running \texttt{python plot.py}; the figure will pop-up in a separate window.
\end{description}

All of the parameters used in \texttt{run\_all.sh} are detailed in the body of the paper. The script \texttt{to\_csv.py} simply gathers the relevant data and applies no transformations to it. The script \texttt{plot.py} is a standard generation of a pyplot figure.

\artout

Task $T_1$ generates a list of JSON files stored in the \texttt{raw/} folder; each of these provides information about each of the experiments run, such as the time spent on each MPS simulation, inner product calculation and MPI communication, as well as memory usage during MPS simulation. The name of each of these JSON files indicates the parameters used to run the experiment, and these also appear in the fields of the dictionary contained in the file. The list of JSON files is provided in the repository within the \texttt{raw.zip} file.
Task $T_2$ generates a CSV file gathering the data from the JSON files that is relevant to generate the figure; this \texttt{results.csv} file is provided in the repository.
Task $T_3$ generates the figure, which appears as a pop-up in a separate window.

\newpage

\newartifact 

\artrel

Artifact $A_2$ generates the data for Figure 6 and plots it.
Figure 6 shows the memory usage of the MPS simulation subroutine for two different parameter configurations.
The experiments use $100$ qubits, using less than $1$ GiB per simualtion. In contrast, a statevector simulator would use $16 \times 2^{100}$ bytes.
As such, it directly supports the claim of contribution $C_{1.1}$, since it would be impossible to simulate these experiments using a statevector simulator.

\artexp

The memory required to simulate the circuit ansatz increases with the number of gates applied.
As discussed in the paper, this increase is exponential in the number of gates applied, but the use of SVD truncation---which causes the sudden drops in memory usage---enables the simulation of our circuit ansatz, supporting contribution $C_{1.1}$.

\arttime

Between 60 and 90 minutes to obtain all data for this artifact, using an NVIDIA A100 GPU.

\artin

\artinpart{Hardware}

Data obtained from experiments on a single NVIDIA A100 GPU from \href{https://docs.nersc.gov/systems/perlmutter/architecture/}{Perlmutter}. Any NVIDIA GPU with Compute Capability +7.0 or high-end CPU should reproduce the same qualitative behaviour. Only a single device is required.

\artinpart{Software}

The source code of the framework presented in our paper is publicly available at \url{https://github.com/PabloAndresCQ/qml-cutensornet}.
Installation instructions, including version dependencies are included in the README file at the root of this repository.
\begin{itemize}
    \item General requirements: Python 3.10,  pytket 1.26.0 (\href{https://tket.quantinuum.com/api-docs/getting_started.html}{url}), pandas 2.2.1 (\href{https://pandas.pydata.org/}{url}), scikit-learn 1.4.1.post1 (\href{https://scikit-learn.org/stable/}{url}), mpi4py 3.1.5 (\href{https://mpi4py.readthedocs.io/en/stable/}{url}).
    \item The GPU backend uses cuQuantum 3.10 (\href{https://docs.nvidia.com/cuda/cuquantum/latest/getting_started/getting_started.html}{url}) and pytket-cutensornet 0.6.0 (\href{https://github.com/CQCL/pytket-cutensornet}{url}).
    \item The CPU backend uses Julia 1.9.4 (\href{https://julialang.org/}{url}), pyJulia 0.6.2 (\href{https://github.com/JuliaPy/pyjulia}{url}) and ITensors 0.3.37 (\href{https://github.com/ITensor/ITensors.jl}{url}).
\end{itemize}

\artinpart{Datasets / Inputs}

Experiments use the Elliptic Bitcoin Dataset, which needs to be downloaded from Kaggle at \url{https://www.kaggle.com/datasets/ellipticco/elliptic-data-set}. Then, run the following command to preprocess the dataset: \texttt{python elliptic\_preproc.py}.

\artinpart{Installation and Deployment}

No code compilation is required. Setup is described in detail in the README file at the root of our repository \url{https://github.com/PabloAndresCQ/qml-cutensornet}. Step by step instructions to reproduce this artifact are provided in the following section.

\artcomp

The workflow is separated in two tasks to be executed in sequence: $T_1 \rightarrow T_2$.

\begin{description}
    \item[$T_1$] Execute the experiments and produce the raw data. Do so by running \texttt{run\_all.sh} from the directory of this artifact (see \textit{Computational Artifacts} section).
    \item[$T_2$] Produce the plot from the data in the output files that were produced by $T_1$. Do so by running \texttt{python plot.py}; the figure will pop-up in a separate window.
\end{description}

All of the parameters used in \texttt{run\_all.sh} are detailed in the body of the paper; in this case the files produced are debugging output logs that indicate the amount of memory in usage after each gate is applied. The script \texttt{plot.py} needs to scan these output logs to gather this information, otherwise it is a standard generation of a pyplot figure.

\artout

Task $T_1$ generates a list of text files stored in the \texttt{raw/} folder; each of these provides a debug log for each of the experiments run. There are two folders \texttt{raw/d6} and \texttt{raw/d12} corresponding to the runs with interaction distance $d=6$ and $d=12$ respectively, as described in the body of the paper.
Task $T_2$ generates the figure, which appears as a pop-up in a separate window.

\newpage

\newartifact 

\artrel

Artifact $A_3$ generates the data for Figure 5 and plots it, as well as printing Table I in the command line.
Figure 5 shows a crossover point between the CPU and GPU backends, beyond which the GPU backend runs faster.
As such, it provides empirical proof of contribution $C_{1.2}$.

\artexp

Figure 5 plots the time it takes to run the expensive computational primitives in our framework for circuit ansatz with increasing qubit interaction distance. We show that, for the given ansatz, beyond $d>9$ is enough for the GPU backend to be faster than the CPU backend, both in MPS simulation and inner product calculation; this our claim in contribution $C_{1.2}$. Table I provides further details on these experiments, displaying the maximum value of the virtual bond dimension and the memory footprint of the resulting MPS after simulation.

\arttime

It took 75 minutes to obtain all data for the GPU runs, using a single NVIDIA A100 GPU.
It took 165 minutes to obtain all data for the CPU runs, using a single AMD EPYC 7763 CPU.

\artin

\artinpart{Hardware}

GPU backend data obtained from experiments on a single NVIDIA A100 GPU from \href{https://docs.nersc.gov/systems/perlmutter/architecture/}{Perlmutter}. Any NVIDIA GPU with Compute Capability +7.0 or high-end CPU should reproduce the same qualitative behaviour. Only a single device is required.
CPU backend data obtained from experiments on a single AMD EPYC 7763 CPU from \href{https://docs.nersc.gov/systems/perlmutter/architecture/}{Perlmutter}.

\artinpart{Software}

The source code of the framework presented in our paper is publicly available at \url{https://github.com/PabloAndresCQ/qml-cutensornet}.
Installation instructions, including version dependencies are included in the README file at the root of this repository.
\begin{itemize}
    \item General requirements: Python 3.10,  pytket 1.26.0 (\href{https://tket.quantinuum.com/api-docs/getting_started.html}{url}), pandas 2.2.1 (\href{https://pandas.pydata.org/}{url}), scikit-learn 1.4.1.post1 (\href{https://scikit-learn.org/stable/}{url}), mpi4py 3.1.5 (\href{https://mpi4py.readthedocs.io/en/stable/}{url}).
    \item The GPU backend uses cuQuantum 3.10 (\href{https://docs.nvidia.com/cuda/cuquantum/latest/getting_started/getting_started.html}{url}) and pytket-cutensornet 0.6.0 (\href{https://github.com/CQCL/pytket-cutensornet}{url}).
    \item The CPU backend uses Julia 1.9.4 (\href{https://julialang.org/}{url}), pyJulia 0.6.2 (\href{https://github.com/JuliaPy/pyjulia}{url}) and ITensors 0.3.37 (\href{https://github.com/ITensor/ITensors.jl}{url}).
\end{itemize}

\artinpart{Datasets / Inputs}

Experiments use the Elliptic Bitcoin Dataset, which needs to be downloaded from Kaggle at \url{https://www.kaggle.com/datasets/ellipticco/elliptic-data-set}. Then, run the following command to preprocess the dataset: \texttt{python elliptic\_preproc.py}.

\artinpart{Installation and Deployment}

No code compilation is required. Setup is described in detail in the README file at the root of our repository \url{https://github.com/PabloAndresCQ/qml-cutensornet}. Step by step instructions to reproduce this artifact are provided in the following section.

\artcomp

The workflow is separated in three tasks to be executed in sequence: $T_1 \rightarrow T_2 \rightarrow T_3$.

\begin{description}
    \item[$T_1$] Execute the experiments and produce the raw data. Do so by running \texttt{run\_all.sh} from the directory of this artifact (see \textit{Computational Artifacts} section).
    \item[$T_2$] Combine the output files that were produced by $T_1$ and generate \texttt{gpu\_results.csv} and \texttt{cpu\_results.csv} files. Do so by running \texttt{python to\_csv.py}.
    \item[$T_3$] Produce the plot from the content of the CSV files generated by $T_2$. Do so by running \texttt{python plot.py}. The figure will pop-up in a separate window, Table I will be printed in the command line.
\end{description}

All of the parameters used in \texttt{run\_all.sh} are detailed in the body of the paper. The script \texttt{to\_csv.py} simply gathers the relevant data and applies no transformations to it. The script \texttt{plot.py} is a standard generation of a pyplot figure.

\artout

Task $T_1$ generates a list of JSON files stored in the \texttt{raw/} folder; each of these provides information about each of the experiments run, such as the time spent on each MPS simulation, inner product calculation and MPI communication, as well as memory usage during MPS simulation. The name of each of these JSON files indicates the parameters used to run the experiment, and these also appear in the fields of the dictionary contained in the file. Experiments run on GPU have their files saved in the \texttt{raw/gpu} folder, and those run on CPUs appear in \texttt{raw/cpu}. The list of JSON files is provided in the repository within the \texttt{raw.zip} file.
Task $T_2$ generates two CSV files gathering the data from the JSON files that is relevant to generate the figure; these \texttt{gpu\_results.csv} and \texttt{cpu\_results.csv} files are provided in the repository.
Task $T_3$ generates the figure, which appears as a pop-up in a separate window, and prints Table I in the command line.

\newpage

\newartifact 

\artrel

Artifact $A_4$ generates the data for Figure 8 and plots it.
Figure 8 shows a breakdown of the total runtime taken by our framework to train a model for different data set sizes, and using different number of parallel processes.
As stated in $C_{1.3}$, we can complete training on a data set of $6400$ entries. Doing so takes three hours, using 32 GPUs. The increase in runtime displayed in the figure fits our predictions, as discussed in the body of the paper and in the section below.

\artexp

Each consecutive bar in Figure 8 increases the size of the data set by a factor of two, as well as using twice as many parallel processors. As expected, the wall clock time for MPS simulation remains constant due to the total runtime increasing linearly with data set size, which is counteracted by the linear increase in parallel processes available. The time spent in calculation of inner products increases by a factor of two between each column due to the number of inner products to be calculated being a square of the data set size, and the number of parallel processors only increasing linearly. More details on the asymptotic complexity of our framework are provided in the body of the paper, and the results in Figure 8 fit our predictions, thus supporting $C_{1.3}$.

\arttime

It took a total of 390 minutes of wall clock time to obtain all data. Multiple NVIDIA A100 GPUs were used (from 2 to 32, depending on the experiment).

\artin

\artinpart{Hardware}

The experiments were run on \href{https://docs.nersc.gov/systems/perlmutter/architecture/}{Perlmutter} GPU nodes with NVIDIA A100 GPUs. As indicated in the figure, for each of the bars a different number of GPUs were used: from 2 to 32. Perlmutter' nodes contain 4 GPUs each, so for the runs with more GPUs multiple nodes had to be employed.

\artinpart{Software}

The source code of the framework presented in our paper is publicly available at \url{https://github.com/PabloAndresCQ/qml-cutensornet}.
Installation instructions, including version dependencies are included in the README file at the root of this repository.
\begin{itemize}
    \item General requirements: Python 3.10,  pytket 1.26.0 (\href{https://tket.quantinuum.com/api-docs/getting_started.html}{url}), pandas 2.2.1 (\href{https://pandas.pydata.org/}{url}), scikit-learn 1.4.1.post1 (\href{https://scikit-learn.org/stable/}{url}), mpi4py 3.1.5 (\href{https://mpi4py.readthedocs.io/en/stable/}{url}).
    \item The GPU backend uses cuQuantum 3.10 (\href{https://docs.nvidia.com/cuda/cuquantum/latest/getting_started/getting_started.html}{url}) and pytket-cutensornet 0.6.0 (\href{https://github.com/CQCL/pytket-cutensornet}{url}).
    \item The CPU backend uses Julia 1.9.4 (\href{https://julialang.org/}{url}), pyJulia 0.6.2 (\href{https://github.com/JuliaPy/pyjulia}{url}) and ITensors 0.3.37 (\href{https://github.com/ITensor/ITensors.jl}{url}).
\end{itemize}

\artinpart{Datasets / Inputs}

Experiments use the Elliptic Bitcoin Dataset, which needs to be downloaded from Kaggle at \url{https://www.kaggle.com/datasets/ellipticco/elliptic-data-set}. Then, run the following command to preprocess the dataset: \texttt{python elliptic\_preproc.py}.

\artinpart{Installation and Deployment}

No code compilation is required. Setup is described in detail in the README file at the root of our repository \url{https://github.com/PabloAndresCQ/qml-cutensornet}. Step by step instructions to reproduce this artifact are provided in the following section.

\artcomp

The workflow is separated in three tasks to be executed in sequence: $T_1 \rightarrow T_2 \rightarrow T_3$.

\begin{description}
    \item[$T_1$] Execute the experiments and produce the raw data. Do so by running \texttt{run\_all.sh} from the directory of this artifact (see \textit{Computational Artifacts} section). This will run multiple slurm jobs, each taking the appropriate number of GPUs. The script is designed to run on Perlmutter using the setup described in the README at the root of the repository. Other computers or setups may require changes to the scripts in the directory \texttt{slurm\_scripts}.
    \item[$T_2$] Combine the output files that were produced by $T_1$ and generate a \texttt{results.csv} file. Do so by running \texttt{python to\_csv.py}.
    \item[$T_3$] Produce the plot from the content of the CSV files generated by $T_2$. Do so by running \texttt{python plot.py}. The figure will pop-up in a separate window.
\end{description}

All of the parameters used in \texttt{run\_all.sh} are detailed in the body of the paper; the variable \texttt{ntr} corresponds to half the data set size, since the latter is comprised of \texttt{ntr} entries labelled ``illicit" and \texttt{ntr} entries labelled ``licit". The script \texttt{to\_csv.py} simply gathers the relevant data and applies no transformations to it. The script \texttt{plot.py} is a standard generation of a pyplot figure.

\artout

Task $T_1$ generates a list of JSON files stored in the \texttt{raw/} folder; each of these provides information about each of the experiments run, such as the wall clock time spent on each MPS simulation, inner product calculation and MPI communication, as well as memory usage during MPS simulation. The name of each of these JSON files indicates the parameters used to run the experiment, and these also appear in the fields of the dictionary contained in the file. The list of JSON files is provided in the repository within the \texttt{raw.zip} file.
Task $T_2$ generates a CSV file gathering the data from the JSON files that is relevant to generate the figure; the \texttt{results.csv} file is provided in the repository.
Task $T_3$ generates the figure, which appears as a pop-up in a separate window.

\newpage

\newartifact 

\artrel

Artifact $A_5$ generates the data for Figure 9 and 10 and plots them.
These figures plot the AUC metric (an indicator of classification performance) of our model for different data set sizes and number of features. Figure 9 shows the AUC metric for the training set, whereas Figure 10 shows the AUC metric for the test set.
As stated in $C_{2.1}$, the AUC metric improves with more features and larger data sets. Some instability is perceived for smaller data sets, which is to be expected, as discussed in the body of the paper.

\artexp

Figure 10 is the more relevant of the two, since it indicates the quality of prediction of our model.
In Figure 10 we see that the best results are achieved with the largest number of features (165) and the largest data set (6400); this corroborates $C_{2.1}$.
For the experiments on dataset of size 6400 we see a steady improvement with the increase in number of features; for smaller data sets the conclusion is less definitive due to overfitting, as discussed in the body of the paper.

\arttime

It took a total of 850 minutes of wall clock time to obtain all data. Multiple NVIDIA A100 GPUs were used; the experiments on data set size $300$ and $1500$ used $4$ GPUs each, the experiments on data set size $6400$ used $32$ GPUs each.

\artin

\artinpart{Hardware}

The experiments were run on \href{https://docs.nersc.gov/systems/perlmutter/architecture/}{Perlmutter} GPU nodes with NVIDIA A100 GPUs. As indicated above, different experiments used different number of GPUs: either 4 or 32. Perlmutter' nodes contain 4 GPUs each, so for the runs with 32 GPUs, 8 different nodes had to be employed.

\artinpart{Software}

The source code of the framework presented in our paper is publicly available at \url{https://github.com/PabloAndresCQ/qml-cutensornet}.
Installation instructions, including version dependencies are included in the README file at the root of this repository.
\begin{itemize}
    \item General requirements: Python 3.10,  pytket 1.26.0 (\href{https://tket.quantinuum.com/api-docs/getting_started.html}{url}), pandas 2.2.1 (\href{https://pandas.pydata.org/}{url}), scikit-learn 1.4.1.post1 (\href{https://scikit-learn.org/stable/}{url}), mpi4py 3.1.5 (\href{https://mpi4py.readthedocs.io/en/stable/}{url}).
    \item The GPU backend uses cuQuantum 3.10 (\href{https://docs.nvidia.com/cuda/cuquantum/latest/getting_started/getting_started.html}{url}) and pytket-cutensornet 0.6.0 (\href{https://github.com/CQCL/pytket-cutensornet}{url}).
    \item The CPU backend uses Julia 1.9.4 (\href{https://julialang.org/}{url}), pyJulia 0.6.2 (\href{https://github.com/JuliaPy/pyjulia}{url}) and ITensors 0.3.37 (\href{https://github.com/ITensor/ITensors.jl}{url}).
\end{itemize}

\artinpart{Datasets / Inputs}

Experiments use the Elliptic Bitcoin Dataset, which needs to be downloaded from Kaggle at \url{https://www.kaggle.com/datasets/ellipticco/elliptic-data-set}. Then, run the following command to preprocess the dataset: \texttt{python elliptic\_preproc.py}.

\artinpart{Installation and Deployment}

No code compilation is required. Setup is described in detail in the README file at the root of our repository \url{https://github.com/PabloAndresCQ/qml-cutensornet}. Step by step instructions to reproduce this artifact are provided in the following section.

\artcomp

The workflow is separated in two tasks to be executed in sequence: $T_1 \rightarrow T_2$.

\begin{description}
    \item[$T_1$] Execute the experiments and produce the raw data. Do so by running \texttt{run\_all.sh} from the directory of this artifact (see \textit{Computational Artifacts} section). This will run multiple slurm jobs, each taking the appropriate number of GPUs. The script is designed to run on Perlmutter using the setup described in the README at the root of the repository. Other computers or setups may require changes to the scripts in the directory \texttt{slurm\_scripts}.
    \item[$T_2$] Produce the plots from the content of files generated by $T_1$. Do so by running \texttt{python plot.py}. The figures will pop-up in a separate window; first Figure 9, then Figure 10.
\end{description}

All of the parameters used in \texttt{run\_all.sh} are detailed in the body of the paper; the variable \texttt{ntr} corresponds to half the data set size, since the latter is comprised of \texttt{ntr} entries labelled ``illicit" and \texttt{ntr} entries labelled ``licit". Each experiment computes its classification performance metrics for different values of the SVM regularization coefficient (from $0.01$ to $4.0$, as discussed in the body of the paper). The script \texttt{plot.py} gathers the AUC metrics and, for each experiment, it picks the best AUC score among the different possible choices of the regularization coefficient; otherwise the script is a standard generation of a pyplot figure.

\artout

Task $T_1$ generates a list of \texttt{.npy} files stored in the \texttt{raw/} folder; each of these provides a list of tuples of the form \texttt{(reg,accuracy, precision, recall, auc)}, where \texttt{reg} is the value of the regularization coefficient, and the other entries are the corresponding classification performance metrics for the given experiment and the chosen regularization coefficient. 
The name of each of these \texttt{.npy} files indicates the parameters used to run the experiment. The list of \texttt{.npy} files is provided in the repository within the \texttt{raw.zip} file.
Task $T_2$ generates the two figures, which appear as a pop-up in a separate window.

\newpage

\newartifact 

\artrel

Artifact $A_6$ generates the data for Table II.
This table displays the standard classification performance metrics of our model on the test data set.
Different experiments are carried out with different values of the qubit interaction distance ($d$) and kernel bandwidth parameter ($\gamma$).
This computational artifact also runs a standard Gaussian kernel SVM to compare against our quantum kernel framework; results are reported in the first row of the table.
The results of Table II support $C_{2.2}$ and $C_{2.3}$, as discussed below.

\artexp

As stated in $C_{2.2}$, Table II shows that our quantum kernel framework outperforms the Gaussian kernel across multiple classification performance metrics.
As stated in $C_{2.3}$, Table II shows that increasing the complexity of the quantum circuit ansatz (model expressivity) does not translate to an improvement in classification performance in our studied dataset.

\arttime

All experiments combined complete in a total of less than 1600 minutes of wall clock time. Each experiment used 4 NVIDIA A100 GPUs.

\artin

\artinpart{Hardware}

The experiments were run on \href{https://docs.nersc.gov/systems/perlmutter/architecture/}{Perlmutter} GPU nodes with NVIDIA A100 GPUs. As indicated above, each experiment uses 4 GPUs and, hence, a single Perlmutter node.

\artinpart{Software}

The source code of the framework presented in our paper is publicly available at \url{https://github.com/PabloAndresCQ/qml-cutensornet}.
Installation instructions, including version dependencies are included in the README file at the root of this repository.
\begin{itemize}
    \item General requirements: Python 3.10,  pytket 1.26.0 (\href{https://tket.quantinuum.com/api-docs/getting_started.html}{url}), pandas 2.2.1 (\href{https://pandas.pydata.org/}{url}), scikit-learn 1.4.1.post1 (\href{https://scikit-learn.org/stable/}{url}), mpi4py 3.1.5 (\href{https://mpi4py.readthedocs.io/en/stable/}{url}).
    \item The GPU backend uses cuQuantum 3.10 (\href{https://docs.nvidia.com/cuda/cuquantum/latest/getting_started/getting_started.html}{url}) and pytket-cutensornet 0.6.0 (\href{https://github.com/CQCL/pytket-cutensornet}{url}).
    \item The CPU backend uses Julia 1.9.4 (\href{https://julialang.org/}{url}), pyJulia 0.6.2 (\href{https://github.com/JuliaPy/pyjulia}{url}) and ITensors 0.3.37 (\href{https://github.com/ITensor/ITensors.jl}{url}).
\end{itemize}

\artinpart{Datasets / Inputs}

Experiments use the Elliptic Bitcoin Dataset, which needs to be downloaded from Kaggle at \url{https://www.kaggle.com/datasets/ellipticco/elliptic-data-set}. Then, run the following command to preprocess the dataset: \texttt{python elliptic\_preproc.py}.

\artinpart{Installation and Deployment}

No code compilation is required. Setup is described in detail in the README file at the root of our repository \url{https://github.com/PabloAndresCQ/qml-cutensornet}. Step by step instructions to reproduce this artifact are provided in the following section.

\artcomp

The workflow is separated in two tasks to be executed in sequence: $T_1 \rightarrow T_2$.

\begin{description}
    \item[$T_1$] Execute the experiments and produce the raw data. Do so by running \texttt{run\_all.sh} from the directory of this artifact (see \textit{Computational Artifacts} section). This will run multiple slurm jobs, each taking the appropriate number of GPUs. The script is designed to run on Perlmutter using the setup described in the README at the root of the repository. Other computers or setups may require changes to the scripts in the directory \texttt{slurm\_scripts}.
    \item[$T_2$] Produce a CSV file from the content of files generated by $T_1$. Do so by running \texttt{python to\_csv.py}. The resulting CSV file contains Table II, which is also printed in the command line.
\end{description}

All of the parameters used in \texttt{run\_all.sh} are detailed in the body of the paper; the variable \texttt{ntr} corresponds to half the data set size ($400/2$), since the latter is comprised of \texttt{ntr} entries labelled ``illicit" and \texttt{ntr} entries labelled ``licit". Each experiment computes its classification performance metrics for different values of the SVM regularization coefficient (from $0.01$ to $4.0$, as discussed in the body of the paper). The script \texttt{to\_csv.py} gathers the classification performance metrics and, for each experiment, it first averages each of the metrics over the six different samples taken on common regularization coefficients; then, it picks the regularization coefficient with highest AUC and reports the average score of each metric for said regularization coefficient, which are displayed in Table II. The same workflow is used for the Gaussian kernel and the quantum kernel.

\artout

Task $T_1$ generates a list of \texttt{.npy} files stored in the \texttt{raw/} folder; each of these provides a list of tuples of the form \texttt{(reg,accuracy, precision, recall, auc)}, where \texttt{reg} is the value of the regularization coefficient, and the other entries are the corresponding classification performance metrics for the given experiment and the chosen regularization coefficient. The files in \texttt{raw/gaussian/} correspond to the results of the Gaussian kernel, whereas those in \texttt{raw/quantum/} correspond to the quantum kernel.
The name of each of these \texttt{.npy} files indicates the parameters used to run the experiment. The list of \texttt{.npy} files is provided in the repository within the \texttt{raw.zip} file.
Task $T_2$ generates the table, which is printed in command line and stored in the \texttt{results.csv} file provided in the repository.

\newpage

\newartifact 

\artrel

Artifact $A_7$ generates the data for Table III.
This table displays the standard classification performance metrics of our model on the test data set.
Different experiments are carried out with different values of circuit depth ($r$), i.e. number of ansatz layers.
The results of Table III support $C_{2.3}$, as discussed below.

\artexp

As stated in $C_{2.3}$, Table III shows that increasing the complexity of the quantum circuit ansatz (model expressivity) does not translate to an improvement in classification performance in our studied dataset.

\arttime

All experiments combined complete in a total of less than 800 minutes of wall clock time. Each experiment used 4 NVIDIA A100 GPUs.

\artin

\artinpart{Hardware}

The experiments were run on \href{https://docs.nersc.gov/systems/perlmutter/architecture/}{Perlmutter} GPU nodes with NVIDIA A100 GPUs. As indicated above, each experiment uses 4 GPUs and, hence, a single Perlmutter node.

\artinpart{Software}

The source code of the framework presented in our paper is publicly available at \url{https://github.com/PabloAndresCQ/qml-cutensornet}.
Installation instructions, including version dependencies are included in the README file at the root of this repository.
\begin{itemize}
    \item General requirements: Python 3.10,  pytket 1.26.0 (\href{https://tket.quantinuum.com/api-docs/getting_started.html}{url}), pandas 2.2.1 (\href{https://pandas.pydata.org/}{url}), scikit-learn 1.4.1.post1 (\href{https://scikit-learn.org/stable/}{url}), mpi4py 3.1.5 (\href{https://mpi4py.readthedocs.io/en/stable/}{url}).
    \item The GPU backend uses cuQuantum 3.10 (\href{https://docs.nvidia.com/cuda/cuquantum/latest/getting_started/getting_started.html}{url}) and pytket-cutensornet 0.6.0 (\href{https://github.com/CQCL/pytket-cutensornet}{url}).
    \item The CPU backend uses Julia 1.9.4 (\href{https://julialang.org/}{url}), pyJulia 0.6.2 (\href{https://github.com/JuliaPy/pyjulia}{url}) and ITensors 0.3.37 (\href{https://github.com/ITensor/ITensors.jl}{url}).
\end{itemize}

\artinpart{Datasets / Inputs}

Experiments use the Elliptic Bitcoin Dataset, which needs to be downloaded from Kaggle at \url{https://www.kaggle.com/datasets/ellipticco/elliptic-data-set}. Then, run the following command to preprocess the dataset: \texttt{python elliptic\_preproc.py}.

\artinpart{Installation and Deployment}

No code compilation is required. Setup is described in detail in the README file at the root of our repository \url{https://github.com/PabloAndresCQ/qml-cutensornet}. Step by step instructions to reproduce this artifact are provided in the following section.

\artcomp

The workflow is separated in two tasks to be executed in sequence: $T_1 \rightarrow T_2$.

\begin{description}
    \item[$T_1$] Execute the experiments and produce the raw data. Do so by running \texttt{run\_all.sh} from the directory of this artifact (see \textit{Computational Artifacts} section). This will run multiple slurm jobs, each taking the appropriate number of GPUs. The script is designed to run on Perlmutter using the setup described in the README at the root of the repository. Other computers or setups may require changes to the scripts in the directory \texttt{slurm\_scripts}.
    \item[$T_2$] Produce a CSV file from the content of files generated by $T_1$. Do so by running \texttt{python to\_csv.py}. The resulting CSV file contains Table III, which is also printed in the command line.
\end{description}

All of the parameters used in \texttt{run\_all.sh} are detailed in the body of the paper; the variable \texttt{ntr} corresponds to half the data set size ($400/2$), since the latter is comprised of \texttt{ntr} entries labelled ``illicit" and \texttt{ntr} entries labelled ``licit". Each experiment computes its classification performance metrics for different values of the SVM regularization coefficient (from $0.01$ to $4.0$, as discussed in the body of the paper). The script \texttt{to\_csv.py} gathers the classification performance metrics and, for each experiment, it first averages each of the metrics over the six different samples taken on common regularization coefficients; then, it picks the regularization coefficient with highest AUC and reports the average score of each metric for said regularization coefficient, which are displayed in Table III.

\artout

Task $T_1$ generates a list of \texttt{.npy} files stored in the \texttt{raw/} folder; each of these provides a list of tuples of the form \texttt{(reg,accuracy, precision, recall, auc)}, where \texttt{reg} is the value of the regularization coefficient, and the other entries are the corresponding classification performance metrics for the given experiment and the chosen regularization coefficient.
The name of each of these \texttt{.npy} files indicates the parameters used to run the experiment. The list of \texttt{.npy} files is provided in the repository within the \texttt{raw.zip} file.
Task $T_2$ generates the table, which is printed in command line and stored in the \texttt{results.csv} file provided in the repository.

\end{document}